\documentclass{appolb}
\usepackage{epsfig}
\usepackage{graphicx}
\begin{document}
% \eqsec  % uncomment this line to get equations numbered by (sec.num)
\title{Heavy quark production at collider energies: \\ some selected topics.%
\thanks{Presented at \textit{Strangeness in Quark Matter 2011}, Sept. 18-24, Cracow, Poland.}%
% you can use '\\' to break lines
}
\author{Antoni Szczurek
\address{
The H. Niewodnicza\'{n}ski Institute of Nuclear Physics\\
Polish Academy of Sciences\\
ul. Radzikowskiego 152, 31-342 Krak\'ow, Poland; \\
Rzesz\'ow University,
ul. Rejtana 16A, 35-959 Rzesz\'ow, Poland.
}
%\and
%Antoni Szczurek
%\address{
%University of Rzesz\'ow, PL-35-959 Rzesz\'ow, Poland\\
%Institute of Nuclear Physics PAN, PL-31-342 Cracow, Poland\\
%antoni.szczurek@ifj.edu.pl}
}
\maketitle
\begin{abstract}
We discuss production of charm quarks and mesons as well as nonphotonic 
electrons in $pp$ scattering at RHIC.
The distributions in rapidity and transverse momentum
of charm and bottom quarks/antiquarks are calculated in 
the $k_t$-factorization approach.
The hadronization of heavy quarks is done by means of 
fenomenological fragmentation functions.
The semileptonic decay functions found by fitting
semileptonic decay data measured by the CLEO and BABAR collaborations 
are used. Good description of the inclusive data at large transverse 
momenta of electrons is obtained and a missing strength 
at small transverse momenta of electrons is found.
%Plausible missing mechanisms are discussed.

Furthermore we discuss kinematical correlations between charged leptons 
from semileptonic decays of open charm/bottom, leptons produced in
the Drell-Yan mechanism as well as some other mechanisms not
included so far in the literature.
When calculating the Drell-Yan processes we include transverse momenta of
$q$ and $\bar q$, using the Kwieci\'nski parton distributions.
Reactions initiated by purely QED $\gamma^*\gamma^*$-fusion in elastic 
and inelastic $pp$ collisions as well as diffractive mechanism of 
exclusive $c \bar c$ production are included.
%The contribution of the later mechanism is rather small.
A good description of the dilepton invariant mass spectrum 
measured by the PHENIX collaboration is achieved. 
Predictions for the dilepton pair transverse 
momentum distribution as well as distribution in azimuthal angle 
between electron and positron are presented.
\end{abstract}


%----------------------------  
\section{Introduction}
%----------------------------

Some time ago the PHENIX and STAR collaborations
have measured transverse momentum distribution
of so-called nonphotonic electrons \cite{STAR_electrons,PHENIX_electrons}.
It is a common wisdom that the dominant contribution
to the nonphotonic electrons/positrons comes from the
semileptonic decays of charm and/or beauty mesons.
Formally such processes can be divided into three subsequent stages.
First $c \bar c$ or $b \bar b$ quarks are produced.
The dominant mechanisms being gluon-gluon fusion
and quark-antiquark annihilation close to the threshold.
Next the heavy quarks/antiquarks fragment into heavy
charmed mesons $D, D^*$ or $B, B^*$. The vector
$D^*$ and $B^*$ mesons decay strongly producing 
(pseudo)scalar $D$ and $B$ mesons.
Finally the heavy pseudoscalar mesons decay 
semileptonically producing electrons/positrons.

%The inclusive heavy quark/antiquark production
%can presently be calculated at
%Fixed-Order plus Next-to-Leading-Log (FONLL) level
%\cite{FONLL}.
%An alternative approach for inclusive 
%heavy quark production is $k_t$-factorization
%\cite{CCH91,CE91,BE01,RSS,BS00,HKSST00,LSZ02}. 
%In this approach emission of gluons 
%(see Fig.\ref{fig:diagram_fusion}) is encoded in 
%so-called unintegrated gluon distributions (UGDFs).
%The latter approach is very efficient for description 
%of $Q \bar Q$ correlations \cite{LS06}.

The hadronization of heavy quarks is usually done
with the help of phenomenological fragmentation functions. 
with parameters adjusted to the production of heavy mesons
in $e^+ e^-$ or $p \bar p$ collisions.

The last ingredient are semileptonic decays of heavy mesons.
%Until recently this component was treated by modeling the decay. 
Only recently the CLEO \cite{CLEO} and BABAR \cite{BABAR} 
collaborations has measured precise spectra of electrons/positrons 
coming from the decays of $D$ and $B$ mesons.
This is done by producing specific resonances: $\Psi(3770)$
which decays into $D$ and $\bar D$ mesons (CLEO) and
$\Upsilon(4S)$ which decays into $B$ and $\bar B$ mesons (BABAR).
In both cases the heavy mesons are almost at rest,
so in practice one measures the meson rest frame
distributions of electrons/positrons.

In this presentation the results have been obtained within
the $k_t$-factorization approach. At rather low
RHIC energies intermediate $x$-values
become relevant.  The Kwiecinski unintegrated gluon
(parton) distributions seem the best suited in this context
\cite{Kwiecinski}.
%We shall use also Ivanov-Nikolaev distributions
%which were fitted to deep-inelastic HERA data 
%including intermediate-$x$ region \cite{IN02}.
We shall use both Peterson \cite{Peterson} and so-called perturbative 
\cite{BCFY95} fragmentation functions. The electron/positron decay 
functions fitted recently \cite{LMS09} to the recent CLEO and BABAR data
will be used.

Recently the PHENIX collaboration has measured dilepton invariant mass
spectrum from $0$ to $8$ GeV in $pp$ collisions at $\sqrt{s}=200$ GeV 
\cite{PHENIX}.
%It is commonly believed that the main contribution
%to the dielectron continuum comes from so-called nonphotonic electrons
%which are produced mainly in semileptonic decays of charm and
%bottom mesons. 
Up to now, production of open charm and bottom was studied only in 
inclusive measurements of charmed mesons \cite{Tevatron_mesons} and 
electrons \cite{electrons} and only inclusive observables were
calculated in pQCD approach \cite{CNV05,LMS09}. 
Such predictions give rather good description of the experimental data,
however, the theoretical uncertainties are quite large.
% which makes the situation somewhat clouded and prevents definite conclusions.

Some time ago we have studied kinematical correlations of $c \bar c$
quarks \cite{LS06}, which is, however, difficult to study experimentally.
High luminosity and in a consequence better statistics at present
colliders gives a new possibility to study not only inclusive
distributions but also correlations between outgoing particles.
%(meson-meson, meson-electron or electron-electron). 
Kinematical correlations constitute an alternative method to pin down 
the cross section for charm and bottom production. 

Below I shall limit to presentation of results obtained in
\cite{LMS09,MSS11}. The original presentation at the conference included
also diffractive processes.
%It gives also a great possibility to separate charm and bottom 
%contributions which has a crucial meaning for understanding 
%the character of heavy quarks interactions with the matter 
%created in high energy nuclear collisions \cite{mischke}.   

%---------------------------
\section{Formalism}
%---------------------------

Let us consider the reaction $h_1 + h_2 \to Q + \bar Q + X$,
where $Q$ and $\bar Q$ are heavy quark and heavy antiquark,
respectively.

%--------------------------------------------------------------------

\begin{figure}[!thb]
\begin{center}
\includegraphics[width=4.5cm]{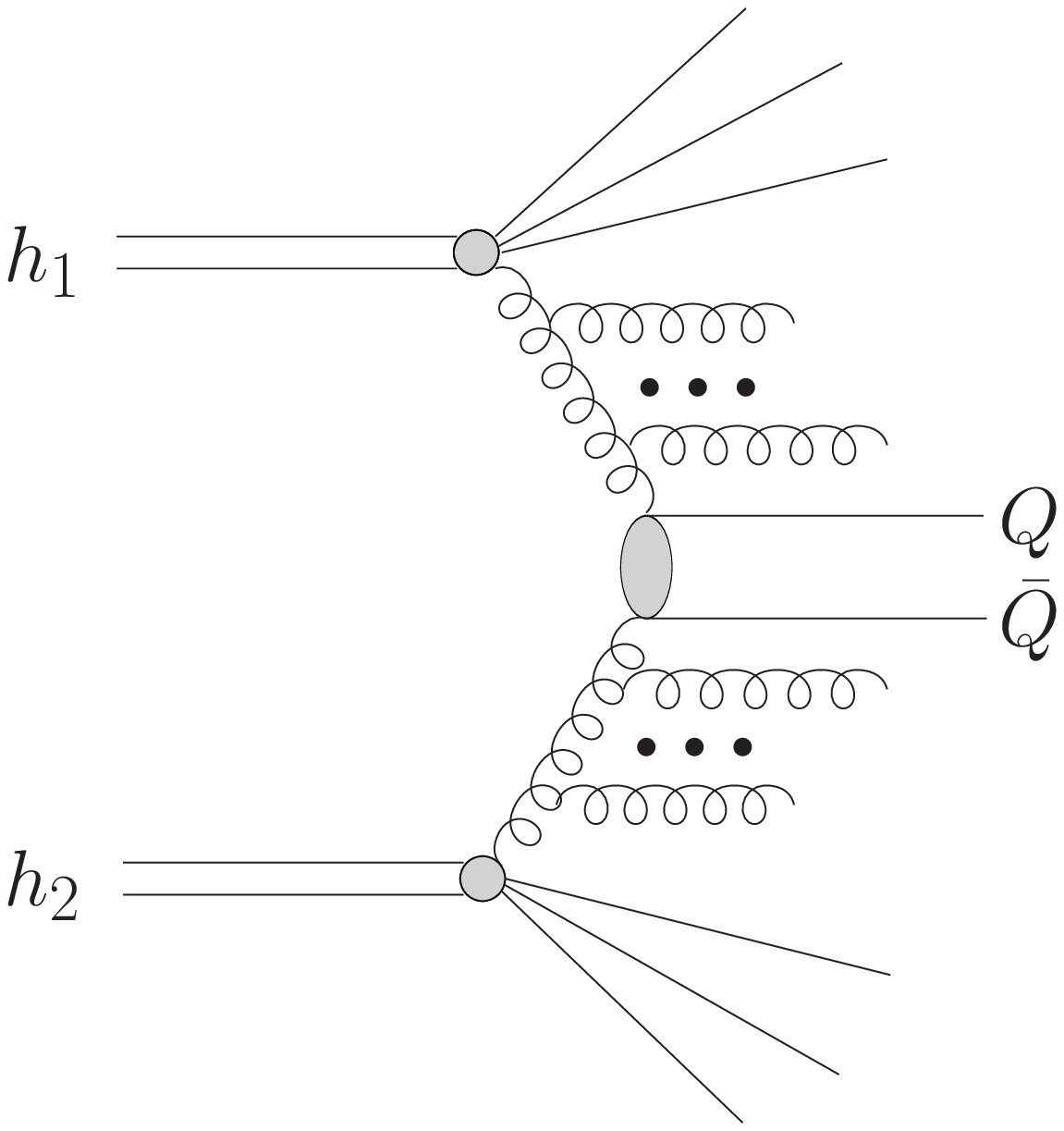}
\includegraphics[width=5.2cm]{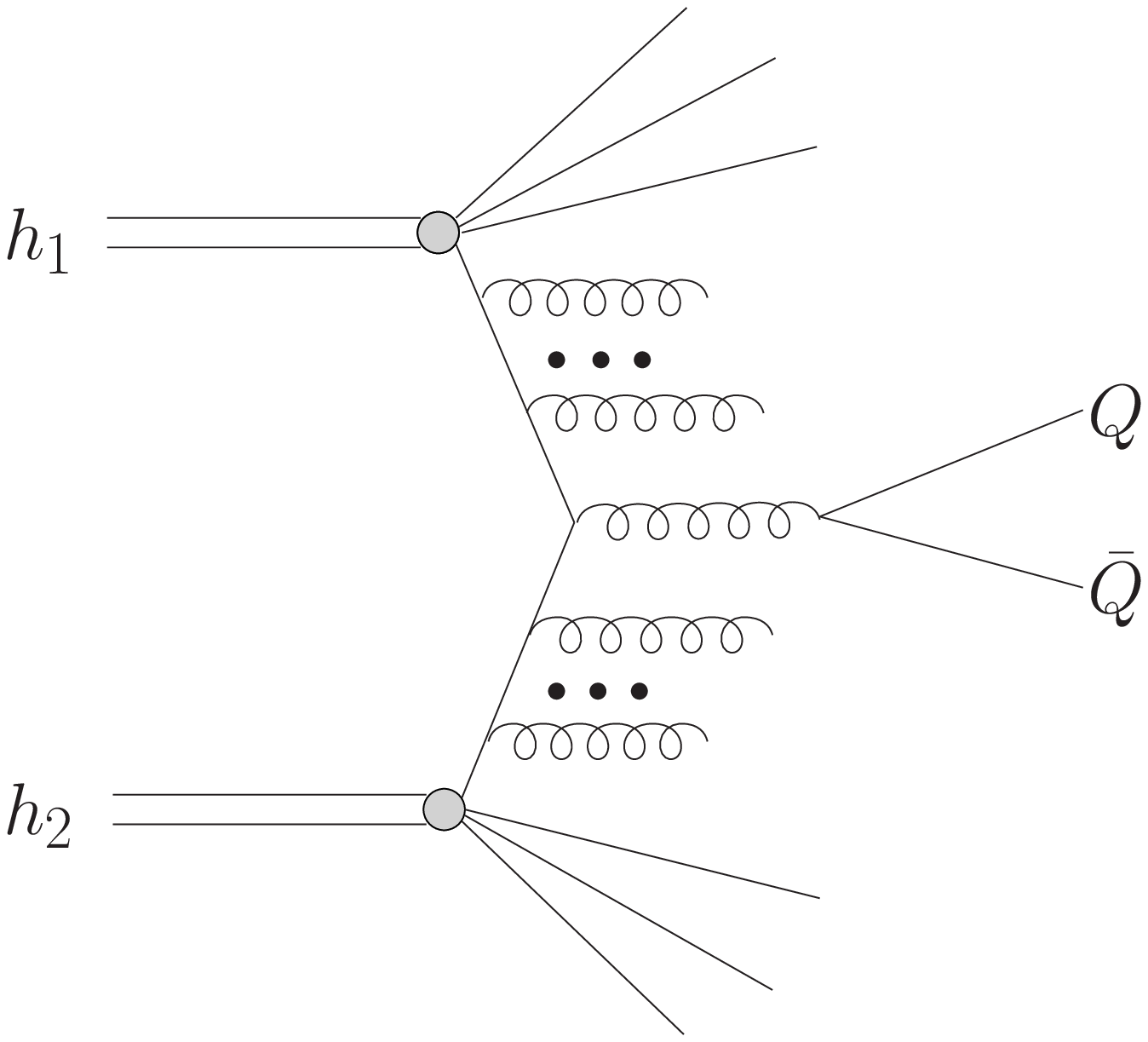}
\caption[*]{
Basic diagrams relevant for gluon-gluon fusion 
(left panel) and quark-antiquark annihilation (right panel)
in our $k_t$-factorization approach.
\label{fig:diagrams}
}
\end{center}
\end{figure}

%---------------------------------------------------------------------

%In the leading-order (LO) approximation within the collinear approach
%the quadruply differential cross section in the rapidity 
%of $Q$ ($y_1$),
%in the rapidity of $\bar Q$ ($y_2$) and the transverse momentum of
%one of them ($p_t$) can be written as
%
%\begin{equation}
%\frac{d \sigma}{d y_1 d y_2 d^2p_t} = \frac{1}{16 \pi^2 {\hat s}^2}
%\sum_{i,j} x_1 p_i(x_1,\mu^2) \; x_2 p_j(x_2,\mu^2) \;
%\overline{|{\cal M}_{ij}|^2} \; .
%\label{LO_collinear}
%\end{equation}
%
%Above, $p_i(x_1,\mu^2)$ and $p_j(x_2,\mu^2)$ are the familiar
%(integrated) parton distributions in hadron $h_1$ and $h_2$, respectively.
%There are two types of the LO $2 \to 2$ subprocesses which enter
%Eq.(\ref{LO_collinear}): $gg \to Q \bar Q$ and $q \bar q \to Q \bar Q$.
%The first mechanism dominates at large energies and the second one
%near the threshold. The parton distributions are evaluated at:
%$x_1 = \frac{m_t}{\sqrt{s}}\left( \exp( y_1) + \exp( y_2) \right)$,
%$x_2 = \frac{m_t}{\sqrt{s}}\left( \exp(-y_1) + \exp(-y_2) \right)$,
%where $m_t = \sqrt{p_t^2 + m_Q^2}$.
%The formulae for matrix element squared averaged over the initial
%and summed over the final spin polarizations can be found e.g. in
%Ref.\cite{BP_book}.

%If one allows for transverse momenta of the initial partons,
%the sum of transverse momenta of the final $Q$ and $\bar Q$ no longer
%cancels.
%Formula (\ref{LO_collinear}) can be easily generalized if
%one allows for the initial parton transverse momenta. Then
In the $k_t$-factorization approach the multiply differential cross
section reads:
\begin{eqnarray}
\frac{d \sigma}{d y_1 d y_2 d^2p_{1,t} d^2p_{2,t}} = \sum_{i,j} \;
\int \frac{d^2 \kappa_{1,t}}{\pi} \frac{d^2 \kappa_{2,t}}{\pi}
\frac{1}{16 \pi^2 (x_1 x_2 s)^2} \; \overline{ | {\cal M}_{ij} |^2}
\nonumber \\  
\delta^{2} \left( \vec{\kappa}_{1,t} + \vec{\kappa}_{2,t} 
                 - \vec{p}_{1,t} - \vec{p}_{2,t} \right) \;
{\cal F}_i(x_1,\kappa_{1,t}^2) \; {\cal F}_j(x_2,\kappa_{2,t}^2) \; ,
\label{LO_kt-factorization}    
\end{eqnarray}
where ${\cal F}_i(x_1,\kappa_{1,t}^2)$ and ${\cal F}_j(x_2,\kappa_{2,t}^2)$
are the so-called unintegrated gluon (parton) distributions.
%\footnote{In this paper we shall use the following convention
%of unintegrated gluon distributions:
%$\int_0^{\mu^2} {\cal F}(x,\kappa^2) d \kappa^2 \sim x g(x,\mu^2)$}.
Leading-order matrix elements for off-shell gluons
\cite{CCH91,CE91,BE01} 
were used.
%The extra integration is over transverse momenta of the initial
%partons.
%The two extra factors $1/\pi$ are due to the integration over
%$d^2 \kappa_{1,t}$ and $d^2 \kappa_{2,t}$ instead over
%$d \kappa_{1,t}^2$ and $d \kappa_{2,t}^2$ as in the conventional
%relation between the unintegrated ($\cal F$) and the integrated ($g$)
%parton distributions. 
The two-dimensional Dirac delta function assures momentum conservation.
The unintegrated parton distributions are evaluated at:\\
$x_1 = \frac{m_{1,t}}{\sqrt{s}}\exp( y_1) 
     + \frac{m_{2,t}}{\sqrt{s}}\exp( y_2)$,
$x_2 = \frac{m_{1,t}}{\sqrt{s}}\exp(-y_1) 
     + \frac{m_{2,t}}{\sqrt{s}}\exp(-y_2)$,
where $m_{i,t} = \sqrt{p_{i,t}^2 + m_Q^2}$.
%In general, the matrix element must be calculated for initial
%off-shell partons. The corresponding formulae for initial gluons
%were calculated in \cite{CCH91,CE91} (see also \cite{BE01}).
%It is easy to check \cite{LS06} that in the limit
%$\kappa_1^2 \to 0$, $\kappa_2^2 \to 0$
%the off-shell matrix elements converge to
%the on-shell ones.

Introducing new variables:
$\vec{Q}_t = \vec{\kappa}_{1,t} + \vec{\kappa}_{2,t} \; , \nonumber \\
\vec{q}_t = \vec{\kappa}_{1,t} - \vec{\kappa}_{2,t}$ 
we can write:
\begin{eqnarray}
\frac{d \sigma_{ij}}{d y_1 d y_2 d^2p_{1,t} d^2p_{2,t}} =
\int d^2 q_t \; \frac{1}{4 \pi^2}
\frac{1}{16 \pi^2 (x_1 x_2 s)^2} \; \overline{ | {\cal M}_{ij} |^2}
\nonumber \\  
{\cal F}_i(x_1,\kappa_{1,t}^2) \; {\cal F}_j(x_2,\kappa_{2,t}^2) \; .
\label{LO_kt-factorization2}    
\end{eqnarray}
This formula is very useful to study correlations between
the produced heavy quark $Q$ and heavy antiquark 
$\bar Q$ \cite{LS06}.

%For example
%
%\begin{eqnarray}
%\frac{d \sigma_{ij}}{d p_{1,t} d p_{2,t}} &=&
%\int d \phi_1 d \phi_2 \; p_{1,t} p_{2,t} \int dy_1 d y_2
%\int d^2 q_t \; \frac{1}{4 \pi^2}
%\frac{1}{16 \pi^2 (x_1 x_2 s)^2} \; \overline{ | {\cal M}_{ij} |^2}
%\nonumber \\  
%&&{\cal F}_i(x_1,\kappa_{1,t}^2) \; {\cal F}_j(x_2,\kappa_{2,t}^2) 
%\nonumber \\
%&=& 4 \pi \; \frac{1}{2} \; \frac{1}{2} \;
% \int d \phi_{-} \; p_{1,t} p_{2,t} \int dy_1 d y_2
%\int d^2 q_t \; \frac{1}{4 \pi^2}
%\frac{1}{16 \pi^2 (x_1 x_2 s)^2} \; \overline{ | {\cal M}_{ij} |^2}
%\nonumber \\  
%&&{\cal F}_i(x_1,\kappa_{1,t}^2) \; {\cal F}_j(x_2,\kappa_{2,t}^2)  \; .
%\label{p1t_p2t_map}
%\end{eqnarray}
%
%In the last equation we have introduced $\phi_{-} \equiv \phi_1 -
%\phi_2$, where $\phi_{-} \in$ (-2$\pi$, 2$\pi$).
%The factor 4 $\pi$ comes from the integration over
%$\phi_{+} \equiv \phi_1 + \phi_2$. The first factor 1/2 comes from
%the jacobian transformation while the second factor 1/2
%takes into account an extra extension of the domain
%when using $\phi_{+}$ and $\phi_{-}$ instead of $\phi_{1}$
%and $\phi_{2}$.

At the Tevatron and LHC energies the contribution of 
the $gg \to Q \bar Q$ subrocess is more than one order 
of magnitude larger than its counterpart for 
the $q \bar q \to Q \bar Q$ subprocess. At RHIC energy
the relative contribution of the $q \bar q$ 
annihilation is somewhat bigger.
Therefore in the following we shall take into account
not only gluon-gluon fusion process
but also the quark-antiquark annihilation mechanism.

%The purely perturbative\footnote{when both UGDFs are generated perturbatively}
%$k_t$-factorization formalism to $h_1 h_2 \to Q \bar Q$ applies if
%$\kappa_{1,t}^2, \kappa_{2,t}^2 > \kappa_0^2$.
%The choice of $\kappa_0^2$ is to a large extent arbitrary.
%In Refs.\cite{RSS} a rather large $\kappa_0^2$ was chosen
%and the space $\kappa_{1,t}^2 \times \kappa_{2,t}^2$ was
%subdivided into four disjoint regions. For example
%the contribution when both $\kappa_{1,t}^2$ and $\kappa_{2,t}^2$
%are small was replaced by the leading-order collinear cross section.
%Such an approach assures that
%$\sigma_{Q \bar Q}^{tot} > \sigma_{Q \bar Q}^{tot}$(collinear LO)
%by construction. \\
%It is rather obvious that the resulting cross section 
%strongly depends on the choice of $\kappa_0^2$
%which makes the procedure a bit arbitrary. 
%Our philosophy here is different. Many models of UGDF 
%in the literature treat the soft region explicitly.
%Therefore we use the $k_t$-factorization formula 
%everywhere on the $\kappa_{1,t}^2 \times \kappa_{2,t}^2$ 
%plane. 

The production of electrons/positrons is a multi-step
process.
The whole procedure of electron/positron production
can be written in the following schematic way:
\begin{equation}
\frac{d \sigma^e}{d y d^2 p} =
\frac{d \sigma^Q}{d y d^2 p} \otimes
D_{Q \to D} \otimes
f_{D \to e} \; ,
\label{whole_procedure}
\end{equation}
where the symbol $\otimes$ denotes a generic convolution.
The first term responsible for production
of heavy quarks/antiquarks.
% is calculated in the $k_t$-factorization approach.
%Some details were already discussed above.
Next step is the process of formation of heavy mesons.
We follow a phenomenological approach and take e.g. Peterson 
\cite{Peterson} and Braaten et al. \cite{BCFY95} fragmentation functions
with parameters from the literature \cite{PDG_new}.
The electron decay function accounts for the proper branching fractions. 
%The latter are known experimentally (see e.g. \cite{PDG,CLEO,BABAR}). 
%These functions can in principle be calculated 
%\cite{Hill,AMP08}. 
%This introduces, however, some model 
%uncertainties and requires inclusion of all final 
%state channels explicitly. An alternative is to use 
%experimental input.
%The decay functions have been measured only recently
%\cite{CLEO,BABAR}. 
%How to use the recent experimental 
%information will be discussed in the next section.

%The hadronization of heavy quarks is usually done
%with the help of fragmentation functions. 
The inclusive distributions of hadrons can be obtained through a
convolution of inclusive distributions of heavy quarks/antiquarks and 
Q $\to$ h fragmentation functions:
\begin{equation}
\frac{d \sigma (y_1, p_{1t}^{H}, y_{2}, p_{2t}^{H}, \phi)}{d y_1 d p_{1t}^{H} d y_{2} d p_{2t}^{H} d \phi}
 \approx
\int \frac{D_{Q \to H}(z_{1})}{z_{1}}\cdot \frac{D_{\bar Q \to \bar H}(z_{2})}{z_{2}}\cdot
\frac{d \sigma (y_1, p_{1t}^{Q}, y_{2}, p_{2t}^{Q}, \phi)}{d y_1 d
  p_{1t}^{Q} d y_{2} d p_{2t}^{Q} d \phi} d z_{1} d z_{2} \; ,
\end{equation}
where: 
{$p_{1t}^{Q} = \frac{p_{1t}^{H}}{z_{1}}$, $p_{2t}^{Q} =
  \frac{p_{2t}^{H}}{z_{2}}$, where
meson longitudinal fractions  $z_{1}, z_{2}\in (0,1)$.
%We have made approximation assuming that $y_{1}, y_{2}, \phi$  are
%unchanged in the fragmentation process.

%There are several models of fragmentation functions in the literature. 
%In our calculations in \cite{LMS09,MSS11} we have used the 
%Peterson fragmentation function \cite{Peterson}. However, to check the
%sensitivity of our results to the choice of the fragmentation model, we 
%have also applied fragmentation functions proposed by Kartvelishvili 
%et al. \cite{kartvel} and Braaten et al. \cite{bcfy}.

%Some time ago the CLEO and BABAR collaborations have measured very precisely
%the spectrum of electrons/positrons coming from
%the weak decays of $D$ and $B$ mesons, respectively \cite{CLEO}.
%These functions can in principle be calculated.
%This introduces, however, some model 
%uncertainties and requires inclusion of all final 
%state channels explicitly. 
We use decay functions fitted recently \cite{LMS09} to the CLEO and BABAR data.
%the experimental input which after renormalizing 
%to experimental branching fractions
%has been used to generate electrons/positrons 
%in a Monte Carlo approach. 
In our approach the electrons (positrons) are
generated isotropically in the heavy meson rest frame.
%The results presented here were obtained with parametrizations of the
%decay functions found in Ref.\cite{LMS09}.

%The whole procedure described above can be written in a schematic way as:
%
%\begin{equation}
%\frac{d \sigma^e}{d y d^2 p} =
%\frac{d \sigma^Q}{d y d^2 p} \otimes
%D_{Q \to h} \otimes
%f_{h \to e} \; ,
%\label{whole_procedure}
%\end{equation}

%---------------------
\section{Results}
%---------------------

%-----------------------------------------------
\subsection{Single electron spectra}
%-----------------------------------------------

Before we start presenting our results for the spectra let us
focus for a moment on the decay functions discussed shortly above.
In Fig.\ref{fig:cleo_and_babar} we show our fit \cite{LMS09} to the
CLEO and BABAR data. The good quality fit of the data will allow us
to obtain reliable predictions for electron/positron single particle spectra.

%--------------------------------------------------------

\begin{figure}[!thb] % 
\begin{center}
\includegraphics[width=5.0cm]{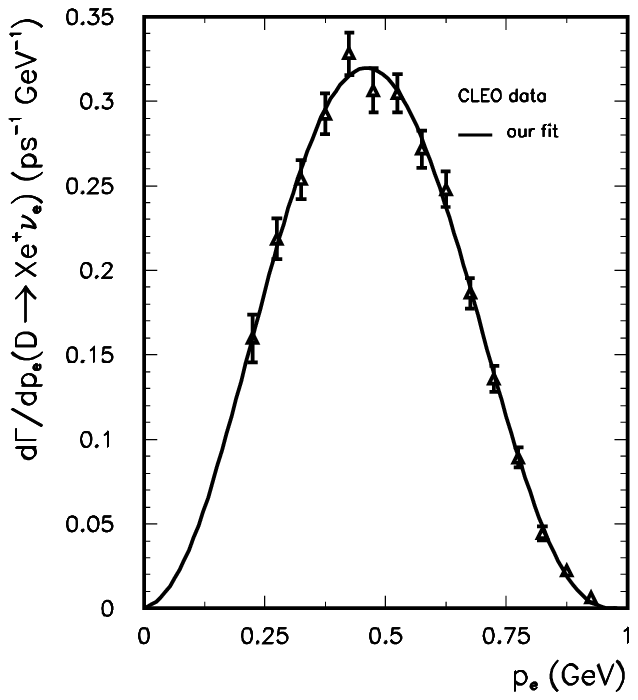}
\includegraphics[width=5.0cm]{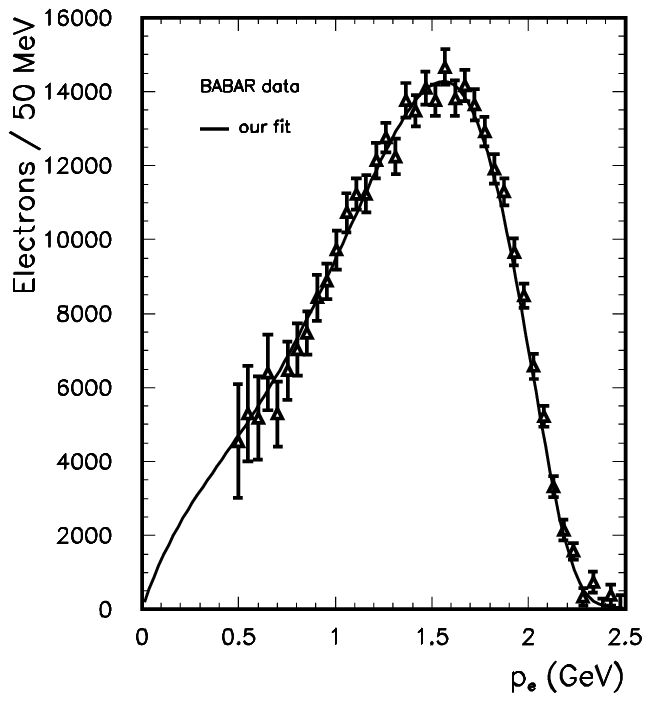}
\end{center}
\caption[*]{
Our fit to the CLEO \cite{CLEO} and BABAR \cite{BABAR}
data.
\label{fig:cleo_and_babar}
}
\end{figure}

%---------------------------------------------------------

%\begin{figure}[!thb]
%\begin{center}
%\includegraphics[width=5.0cm]{fig3a.eps}
%\includegraphics[width=5.0cm]{fig3b.eps}
%\includegraphics[width=5.0cm]{fig3c.eps}
%\caption[*]{
%Two-dimensional distributions in rapidity and transverse
%momentum for charm quark/antiquark, D mesons and
%electrons/positrons.
%\label{fig:c_to_D_to_e}
%}
%\end{center}
%\end{figure}

%--------------------------------------------------------

Now we shall concentrate on transverse momentum distribution
of electrons/positrons measured recently by
the STAR and PHENIX collaborations at RHIC 
\cite{STAR_electrons,PHENIX_electrons}.
In Fig.\ref{fig:dsig_dpt_kwiecinski1}, as an example
%Fig.\ref{fig:dsig_dpt_kwiecinski2} and
%Fig.\ref{fig:dsig_dpt_kwiecinski3} 
we show results obtained with the Kwieci\'nski UPDFs \cite{Kwiecinski}. 
In Ref.\cite{LMS09} we have discussed in addition other UGDFs.
Uncertainties due to different combinations of factorization and 
renormalization scales as well as due to different choices
fragmentation functions
are shown in Fig.\ref{fig:dsig_dpt_kwiecinski_scale_uncertainty_band}
In these calculations we have included both
gluon-gluon fusion as well as quark-antiquark annihilation.
In the last case we use matrix elements with on-shell
formula but for off-shell kinematics (the discussion
of this point can be found in our earlier paper
\cite{LS06}).
%Our calculation is compared to recent experimental
%data of the PHENIX \cite{PHENIX_electrons} and 
%STAR \cite{STAR_electrons} collaborations.
In Ref.\cite{LMS09} we have discussed also uncertainties due to the choice
of quark masses.

%------------------------------------------------------------------
\begin{figure}[!thb] % 
\begin{center}
\includegraphics[width=5cm]{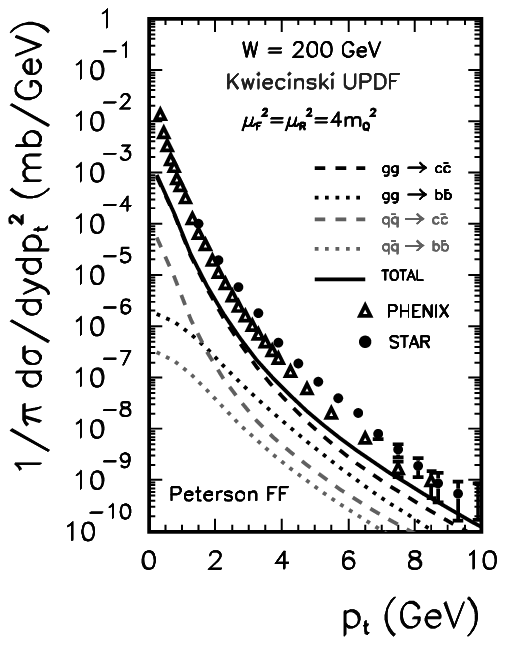}
\includegraphics[width=5cm]{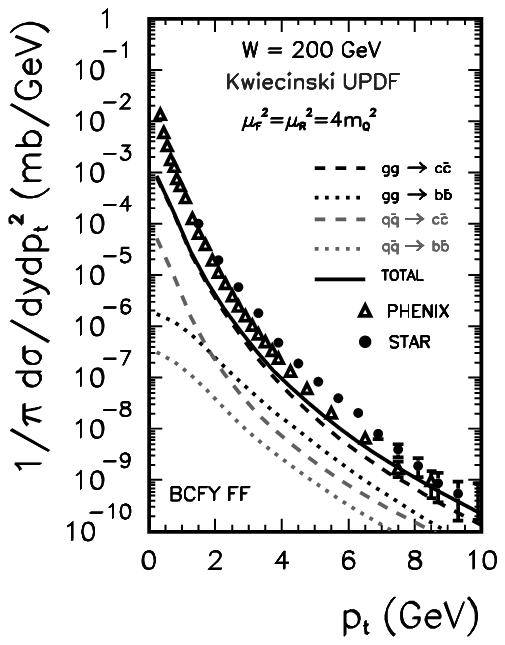}
\end{center}
\caption[*]{
Transverse momentum distribution of electrons/positrons
obtained with the Kwieci\'nski UPDFs. Different combinations
of factorization and renormalization scales are used.
We show separately contributions of the gluon-gluon
fusion (black) and quark-antiquark annihilation (grey).
On the left side results with the Peterson
fragmentation functions and on the right side with
BCFY fragmentation functions. 
\label{fig:dsig_dpt_kwiecinski1}
}
\end{figure}

%--------------------------------------------------------------------

%\begin{figure}[!thb] % 
%\begin{center}
%\includegraphics[width=6cm]{fig5a.eps}
%\includegraphics[width=6cm]{fig5b.eps}
%\includegraphics[width=6cm]{fig5c.eps}
%\includegraphics[width=6cm]{fig5d.eps}

%\caption[*]{
%The same as in the previous section but with different 
%choices of factorization/renormalization scales.
%\label{fig:dsig_dpt_kwiecinski2}
%}
%\end{center}
%\end{figure}

%--------------------------------------------------------------------

%\begin{figure}[!thb] % 
%\begin{center}
%\includegraphics[width=6cm]{fig6a.eps}
%\includegraphics[width=6cm]{fig6b.eps}
%\includegraphics[width=6cm]{fig6c.eps}
%\includegraphics[width=6cm]{fig6d.eps}

%\caption[*]{
%The same as in the previous section but with different 
%choices of factorization/renormalization scales.
%\label{fig:dsig_dpt_kwiecinski3}
%}
%\end{center}
%\end{figure}

%-------------------------------------------------------

Study of nonphotonic $e^{\pm}$ and hadron correlations
allows to "extract" a fractional contribution
of the bottom mesons $B / (D + B)$ as a function
of electron/positron transverse momentum 
\cite{STAR_B_to_DB}. Recently the STAR collaboration
has extended the measurement of the relative
$B$ contribution to electron/positron transverse momenta
$\sim$ 10 GeV \cite{Mischke08a}.
In Fig.\ref{fig:B_fraction_Kwiecinski_scales} 
%and \ref{fig:B_fraction_IN}
(Kwieci\'nski UPDFs) shown are results 
for different scales and different fragmentation functions.
There is a strong dependence on the factorization
and renormalization scale in the case of the Kwieci\'nski
unintegrated gluon/parton distributions.
A slightly better agreement is obtained with the Peterson
fragmentation functions. 

%For Ivanov-Nikolaev gluon
%distribution we show only dependence on fragmentation
%functions.

%-------------------------------------------------------

\begin{figure}[!thb] % 
\begin{center}
\includegraphics[width=5.0cm]{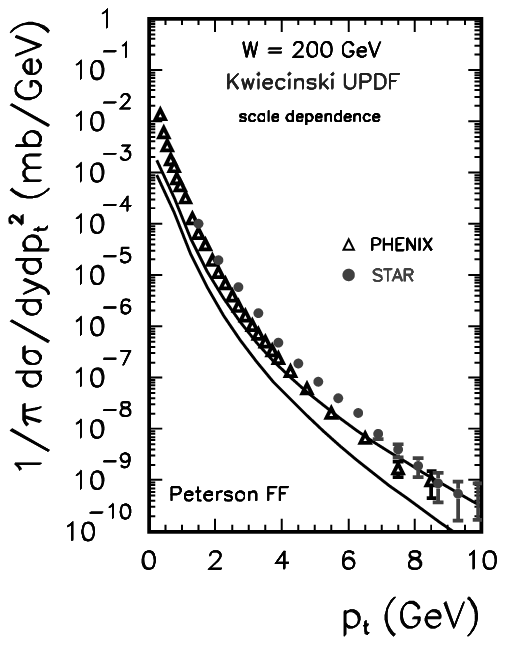}
\includegraphics[width=5.0cm]{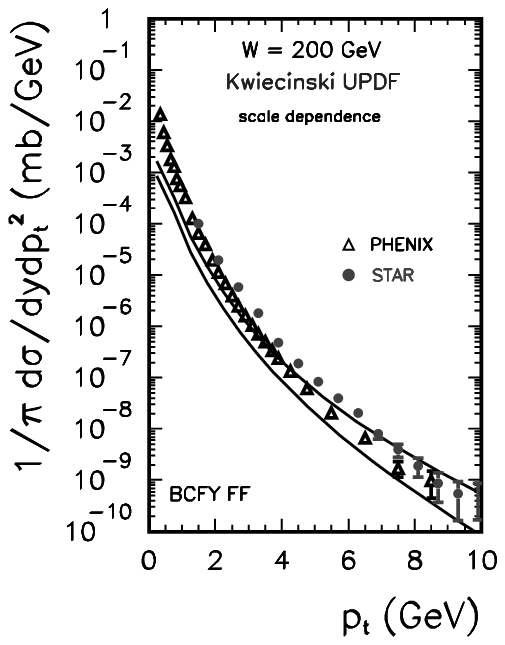}
\end{center}
\caption[*]{
Factorization and renormalization uncertainty band of 
our $k_t$-factorization calculation 
with unintegrated Kwieci\'nski gluon, quark and antiquark
distributions for the Peterson fragmentation function 
(left panel) and BCFY fragmentation function (right panel).
The open triangles represent the PHENIX collaboration data
and the solid circles the STAR collaboration data.
\label{fig:dsig_dpt_kwiecinski_scale_uncertainty_band}
}
\end{figure}

%--------------------------------------------------------------------

%\begin{figure}[!thb] % 
%\begin{center}
%\includegraphics[width=5.0cm]{fig8a.eps}
%\includegraphics[width=5.0cm]{fig8b.eps}
%\end{center}
%\caption[*]{
%An example of quark mass uncertainty band of our 
%$k_t$-factorization calculation  (both D and B decays)
%with unintegrated Kwieci\'nski gluon, quark and antiquark
%distributions for Peterson fragmentation function 
%(left panel).
%The open triangles represent the PHENIX collaboration data
%and the solid circles the STAR collaboration data.
%In the right panel we show the uncertainties separately
%for charm and bottom mesons
%\label{fig:dsig_dpt_kwiecinski_mass_uncertainty_band}
%}
%\end{figure}

%---------------------------------------------------------

%\begin{figure}[!thb] % 
%\begin{center}
%\includegraphics[width=6.0cm]{fig9a.eps}
%\includegraphics[width=6.0cm]{fig9b.eps}
%\caption[*]{
%Transverse momentum distributions of electrons/positrons
%obtained with Ivanov-Nikolaev UGDF and Peterson 
%(left panel) and BCFY (right panel) fragmentation 
%functions.
%\label{fig:dsig_dpt_ivanov_nikolaev}
%}
%\end{center}
%\end{figure}

%=======================================================

\begin{figure}[!thb] % 
\begin{center}
\includegraphics[width=5.0cm]{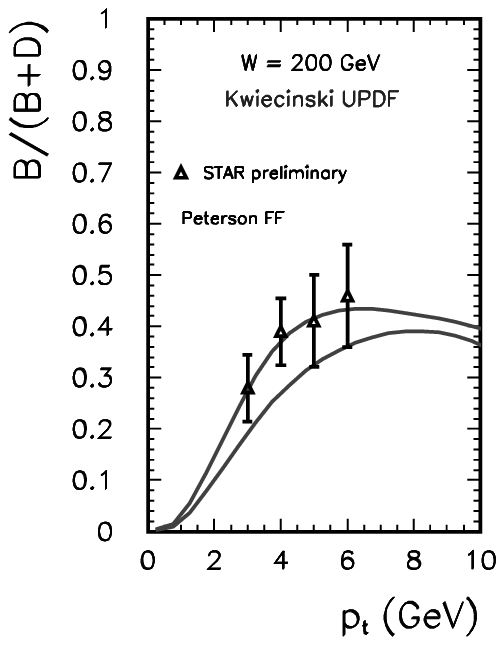}
\includegraphics[width=5.0cm]{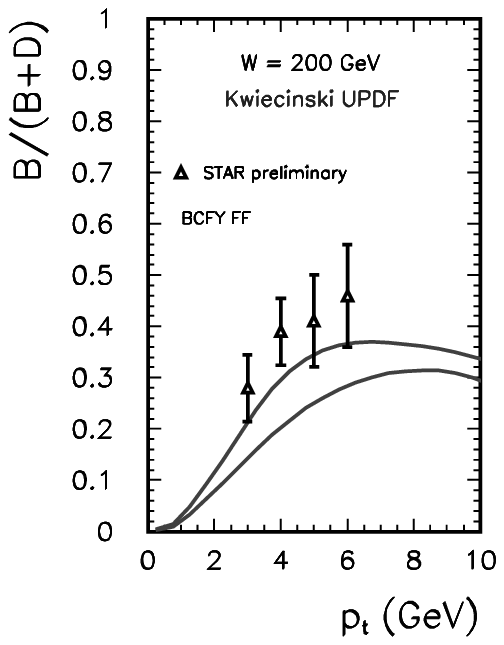}
\end{center}
\caption[*]{
The fraction of the B decays for the Kwieci\'nski
UPDFs. The uncertainty band due to the choice of the scales
is shown for the Peterson (left) and Braaten et al. (right)
fragmentation functions. Both gluon-gluon fusion as well
as quark-antiquark annihilation are included in this 
calculation.
\label{fig:B_fraction_Kwiecinski_scales}
}
\end{figure}

%=======================================================

%\begin{figure}[!thb] % 
%\begin{center}
%\includegraphics[width=6.0cm]{fig11.eps}
%\caption[*]{
%The fraction of the B decays for the Ivanov-Nikolaev
%UGDF for Peterson (solid) and BCFY (dashed).
%Here only gluon-gluon fusion is included as explained
%in the text.
%\label{fig:B_fraction_IN}
%}
%\end{center}
%\end{figure}

%=======================================================

%\begin{figure}[!thb] % 
%\begin{center}
%\includegraphics[width=6.0cm]{fig12.eps}
%\caption[*]{
%The fraction of the B decays for the Kwieci\'nski
%UPDFs. The uncertainty band due to the choice of the quark
%masses is shown for Peterson fragmentation function. 
%Both gluon-gluon fusion as well
%as quark-antiquark annihilation are included in this 
%calculation.
%\label{fig:B_fraction_Kwiecinski_masses}
%}
%\end{center}
%\end{figure}

%=======================================================

%\begin{figure}[!thb] % 
%\begin{center}
%\includegraphics[width=6.0cm]{fig13a.eps}
%\includegraphics[width=6.0cm]{fig13b.eps}
%\caption[*]{
%Uncertainty bands for parameters
%for Peterson (left panel) and BCFY (right panel) 
%fragmentation functions.
%\label{fig:dsig_dpt_uncertainty_parameters}
%}
%\end{center}
%\end{figure}

%=======================================================

%----------------------------------------------------------
\subsection{Electron-positron correlations}
%----------------------------------------------------------

When calculating correlation observables we have included also 
processes shown in Fig.\ref{fig:qed} and Fig.\ref{fig:CED_mechanism}.
The photon-photon induced processes were first included in
Ref. \cite{MSS11}.
The central exclusive diffractive process shown in 
Fig.\ref{fig:CED_mechanism} was first proposed in Ref.\cite{MPS2010}.
%The contributions of the processes will be shown also here.
 
%--------------------------------------------------------------------------
\begin{figure}[!h]
 \includegraphics[width=3.0cm]{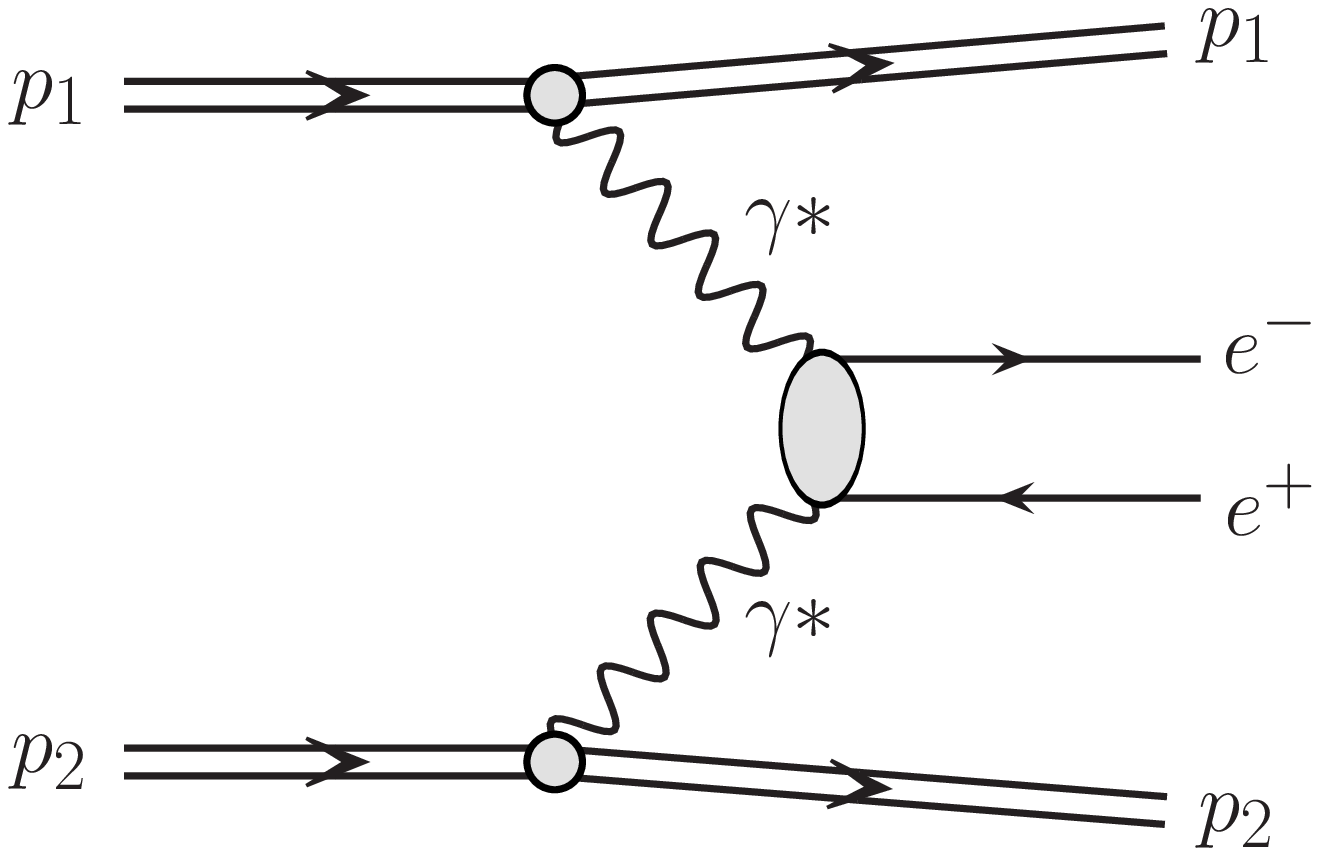}
 \includegraphics[width=3.0cm]{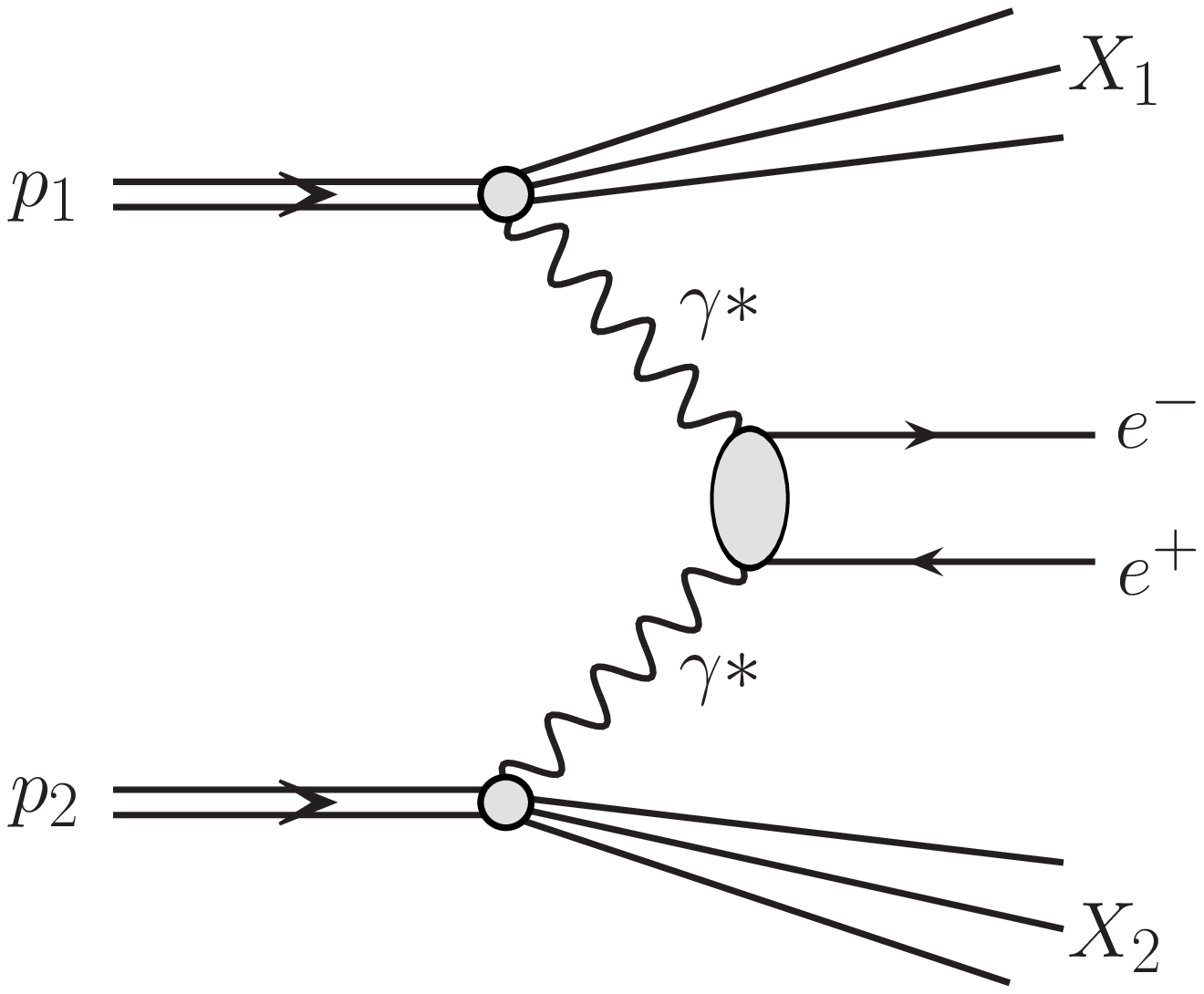}
 \includegraphics[width=3.0cm]{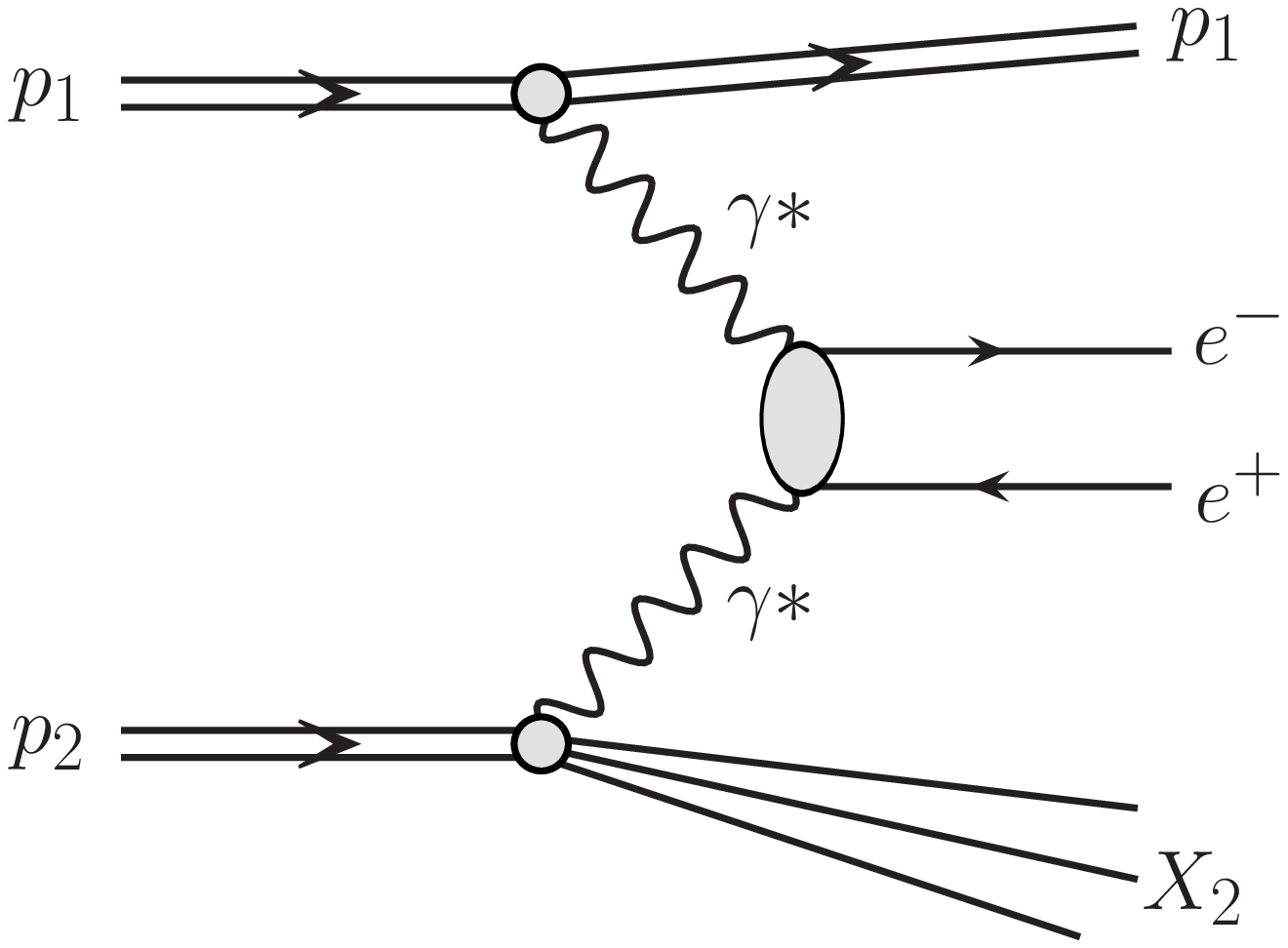}
 \includegraphics[width=3.0cm]{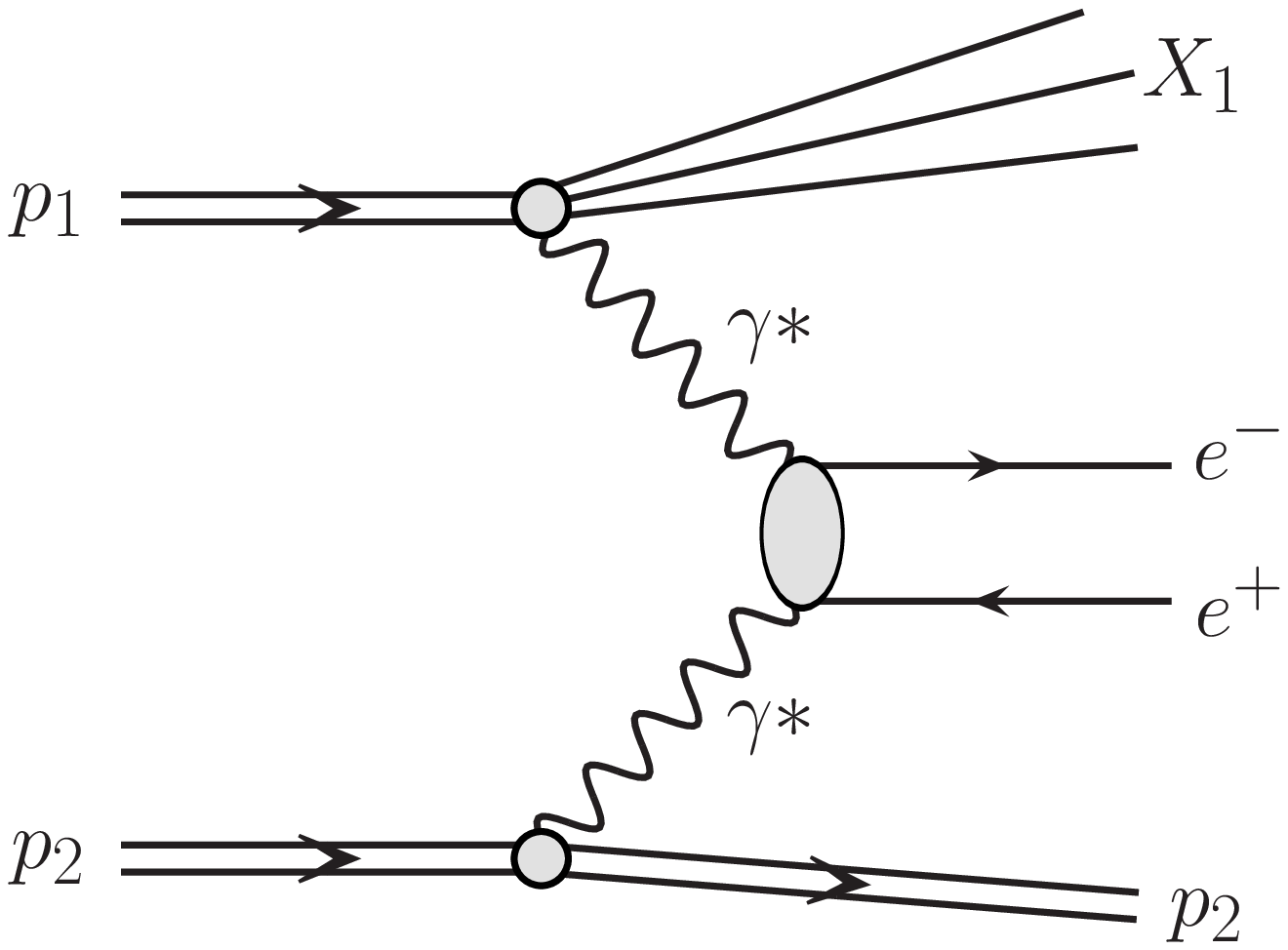}
   \caption{
 \small Diagrammatic representation of processes initiated by
photon-photon subprocesses: double-elastic, 
double-inelastic, inelastic-elastic
and elastic-inelastic.
}
 \label{fig:qed}
\end{figure}
%--------------------------------------------------------------------------

%-------------------------------------------------------------
\begin{figure}[h!]
\begin{center}
\includegraphics[width=5.0cm]{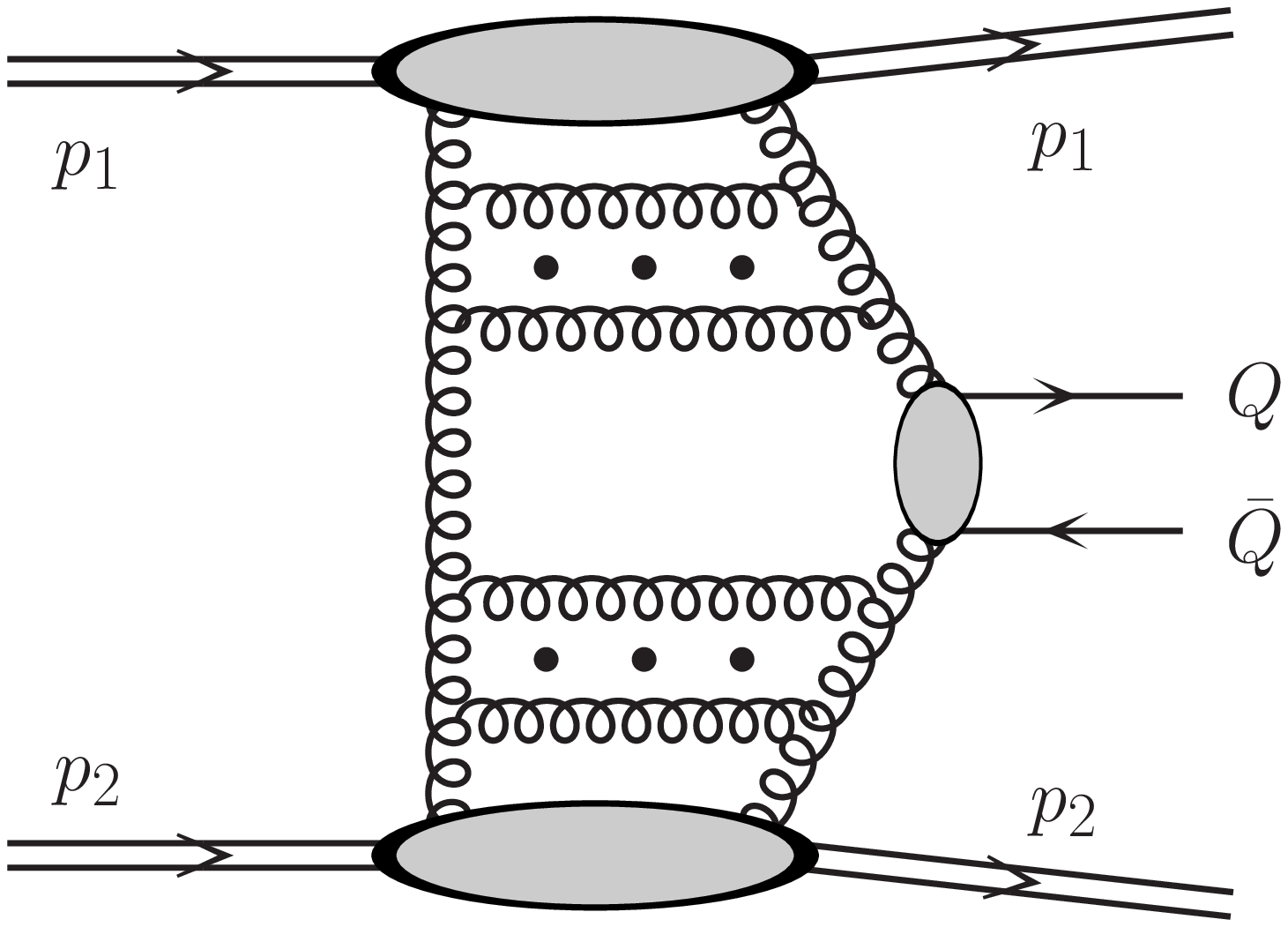}
\end{center}
\caption{\small The mechanism of exclusive double-diffractive
production of open charm.}
\label{fig:CED_mechanism}
\end{figure}
%-------------------------------------------------------------

In Fig.\ref{fig:dsig_dM} we show $e^+  e^-$ invariant mass distributions
calculated with the Kwiecinski (left) and KMR (right) UGDFs. One can
clearly see that both the Kwiecinski and KMR \cite{KMRupdf} UGDFs give
fairly good description of the data for $M_{e^{+}e^{-}} >$ 3 GeV. At small 
invariant masses the Kwiecinski UGDF underestimates the PHENIX data and 
the KMR UGDF starts to overestimate the data points below 
$M_{e^{+}e^{-}}$ = 2 GeV.

%--------------------------------------------------------------------------
\begin{figure}[!h]
\begin{center}
\includegraphics[width=5.0cm]{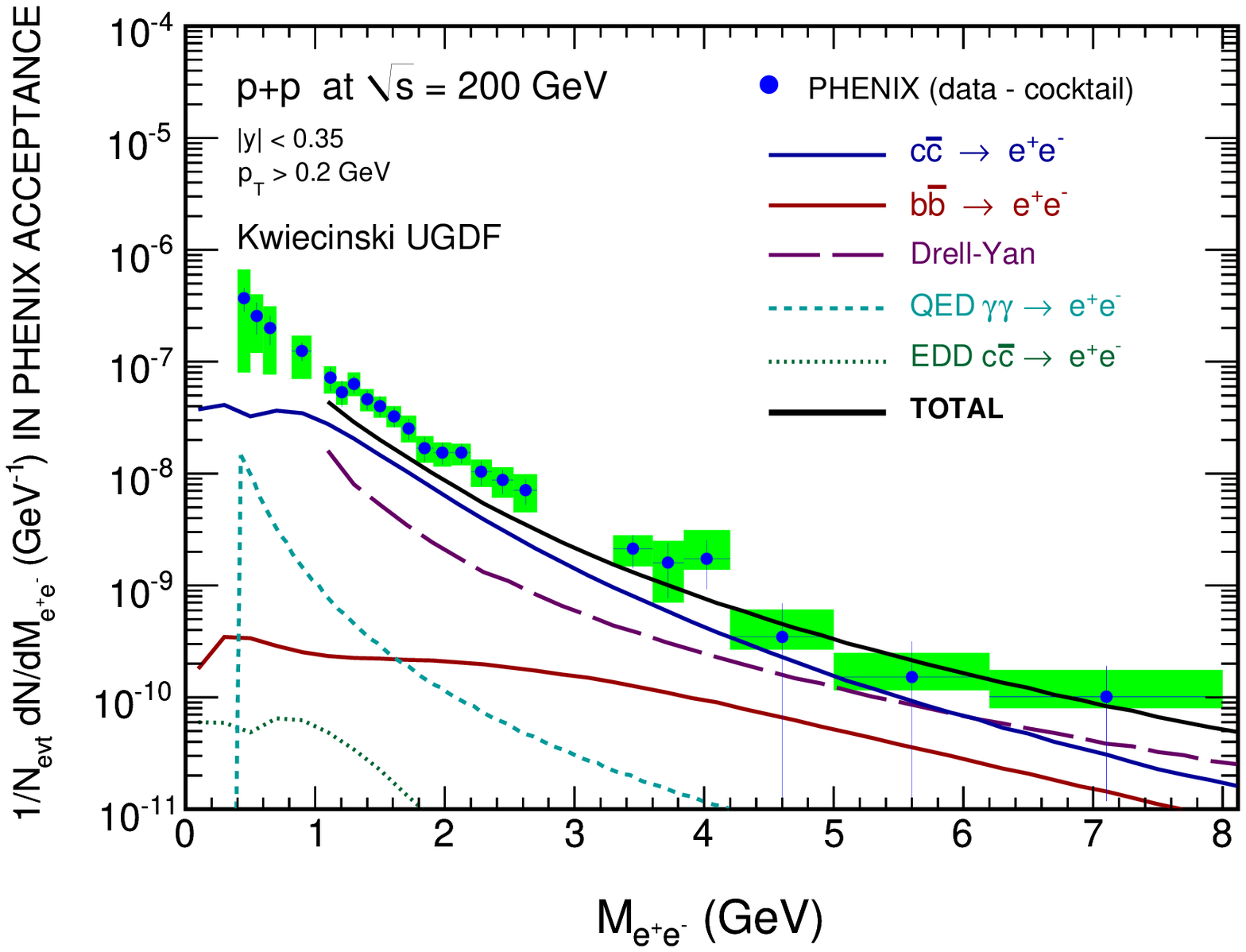}
\includegraphics[width=5.0cm]{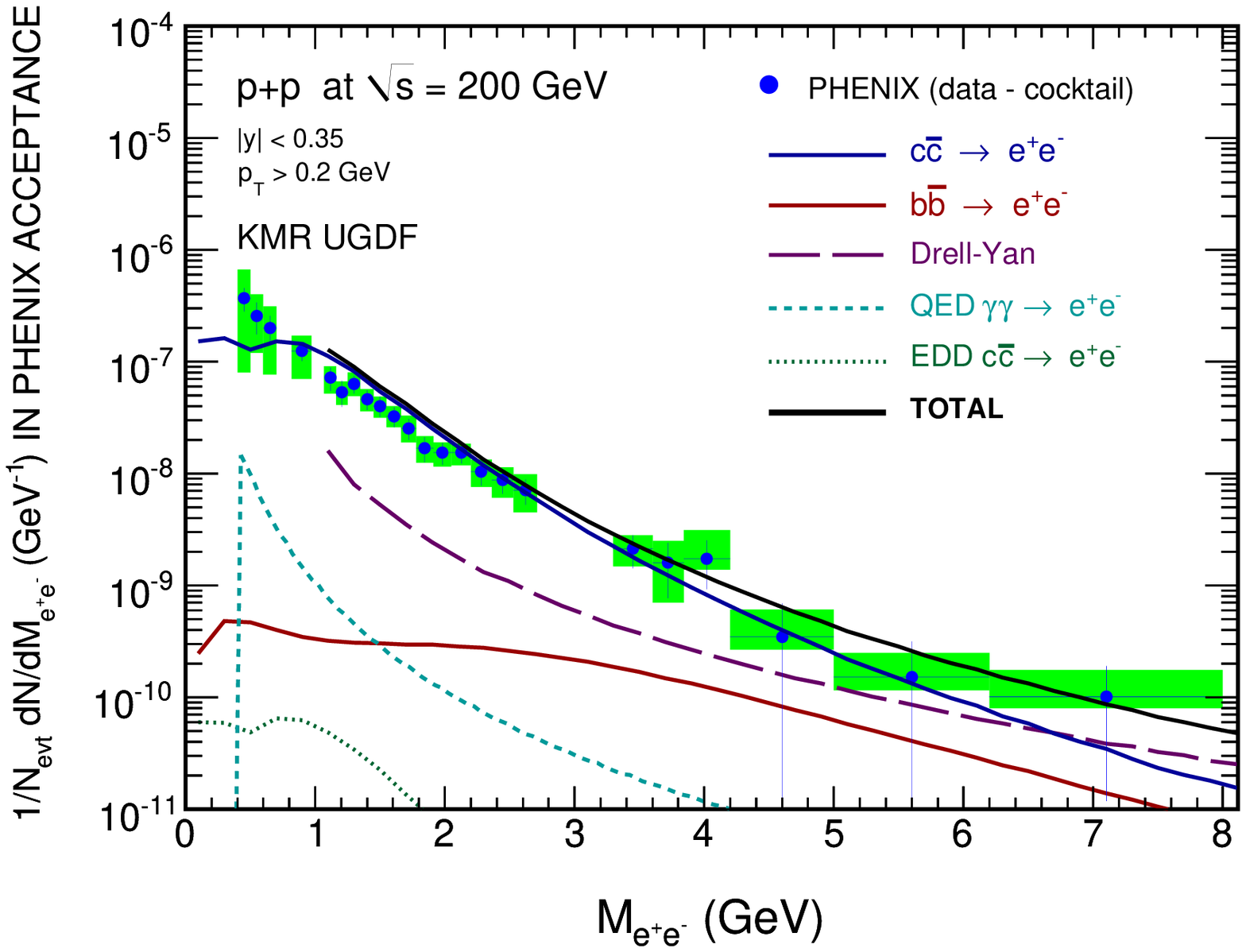}
\end{center}
   \caption{
 \small Dielectron invariant mass distribution for $pp$
collisions at $\sqrt{s}$ = 200 GeV for the Kwieci\'nski (left) and KMR
(right) UGDFs. Different contributions are
shown separately: semileptonic decay of charm by the blue solid line, 
semileptonic decay of bottom by the red solid line, Drell-Yan mechanism by
the long dashed line, gamma-gamma processes by the blue dashed line and 
the central diffractive contribution by the green dotted line.
In this calculation we have included azimuthal angle acceptance of the
PHENIX detector \cite{PHENIX}.
}
 \label{fig:dsig_dM}
\end{figure} 
%-----------------------------------------------------------------------------

%--------------------------------------------------------------------------
%\begin{figure}[!h]
%\begin{center}
% \includegraphics[width=7.0cm]{fig1a-kmr.eps}
%\end{center}
%   \caption{
% \small The same as in Fig.\ref{fig:sources-kwiec} but open charm and bottom components are calculated using
% KMR UGDFs.
%}
% \label{fig:sources-kmr}
%\end{figure} 
%---------------------------------------------------------------------------

%--------------------------------------------------------------------------
%\begin{figure}[!h]
%\begin{center}
% \includegraphics[width=7.0cm]{fig1a-kutak.eps}
%\end{center}
%   \caption{
% \small The same as in Fig.\ref{fig:sources-kwiec} but open charm and bottom components are calculated using
% Kutak-Stasto UGDFs.
%}
% \label{fig:sources-kutak}
%\end{figure} 
%-----------------------------------------------------------------------------

%--------------------------------------------------------------------
%\begin{figure}[!h]
% \includegraphics[width=7.0cm]{fig_qed.eps}
%   \caption{ Contributions of photon-induced mechanisms to
%the dielectron invariant mass distributions. The PHENIX detector
%limitations are given in the upper-left corner.
%}
% \label{fig:gamma_sources}
%\end{figure}
%----------------------------------------------------------------------

In Fig.\ref{fig:uncertainties} we show uncertainties related to
the contribution of semileptonic decays. Complementary the left panel 
presents uncertainties due to the factorization scale variation as described
in the figure caption.
The right panel shows uncertainties due to the 
modification of the heavy quark masses.
% ($m_c \in$ (1.25 GeV, 1.75 GeV) and $m_b \in$ (4.5 GeV, 5 GeV)).

%--------------------------------------------------------------------------
\begin{figure}[!h]
\begin{center}
 \includegraphics[width=5.0cm]{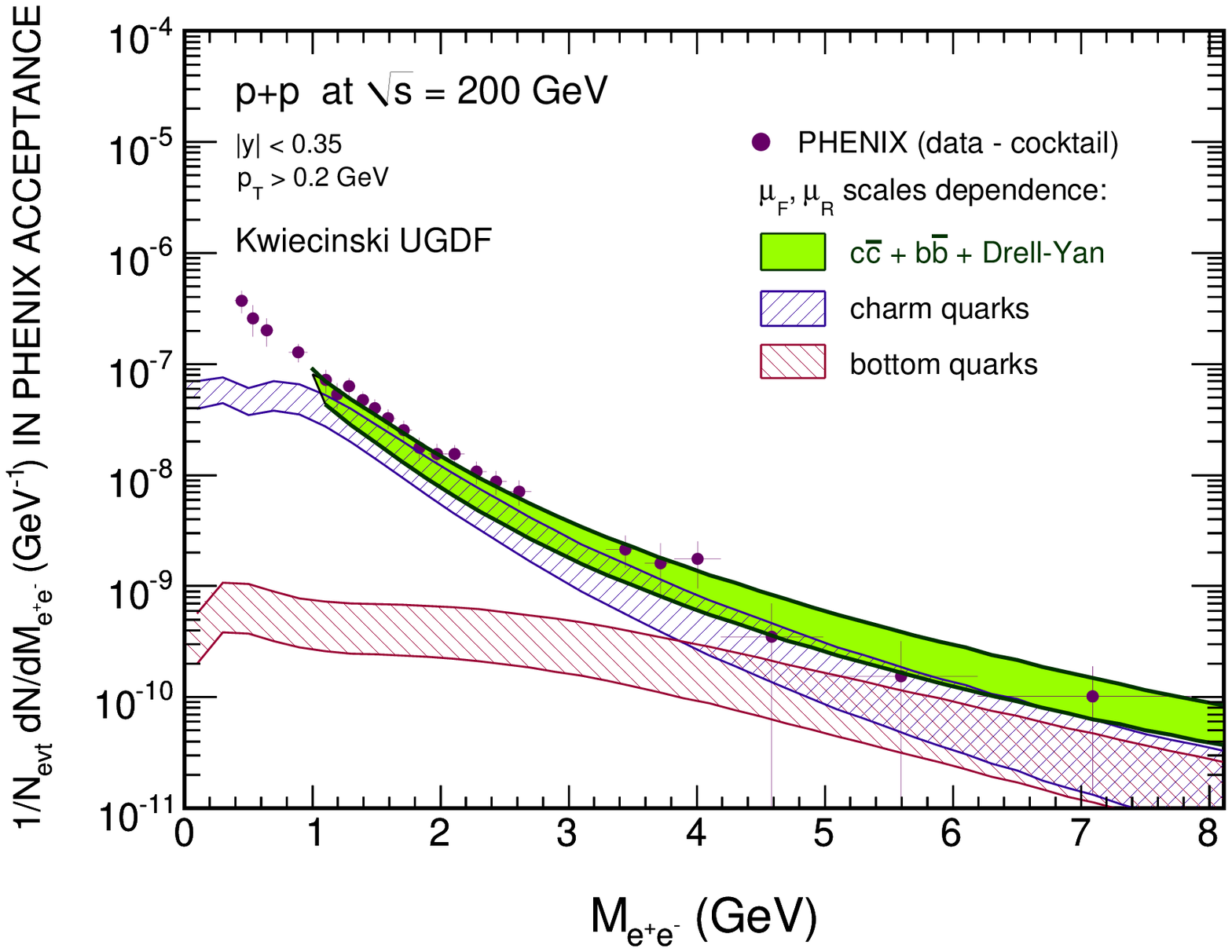}
 \includegraphics[width=5.0cm]{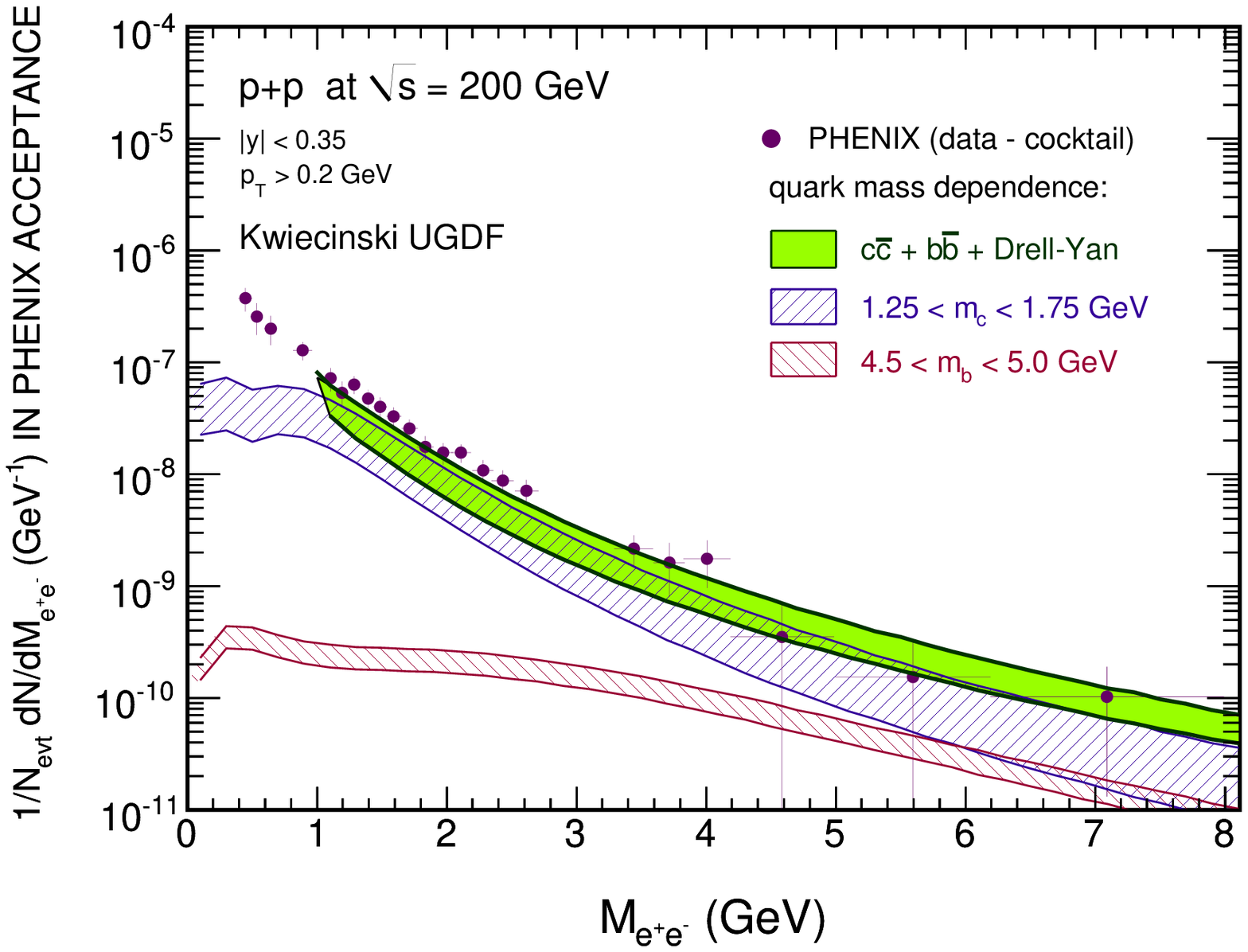}
\end{center}
   \caption{
 \small The uncertainties of theoretical calculations.
The left panel shows the factorization scale uncertainties, 
the lower curve corresponds to $\mu_F^2, \mu_R^2 = m_{1,t}^2 + m_{2,t}^2$ and 
the upper curve to $\mu_R^2 = k_t^2$, $\mu_F^2 = 4 m_Q^2$, where $k_t$ 
is gluon transverse momentum.
The right panel shows the quark mass uncertainties as indicated
in the figure.
}
 \label{fig:uncertainties}
\end{figure}
%-----------------------------------------------------------------------------

%--------------------------------------------------------------------------
%\begin{figure}[!h]
%\begin{center}
%\includegraphics[width=6.0cm]{different_FF.eps}
%\end{center}
%   \caption{
% \small The uncertainties of theoretical calculations of open charm and 
%bottom related to the choice of the fragmentation functions.
%}
% \label{fig:fragmentation}
%\end{figure} 
%-----------------------------------------------------------------------------

%--------------------------------------------------------------------------
%\begin{figure}[!h]
% \includegraphics[width=5.0cm]{quarks.eps}
% \includegraphics[width=5.0cm]{mesons.eps}
% \includegraphics[width=5.0cm]{leptons.eps}
%   \caption{
% \small Two-dimensional distribution in transverse momenta of
%$c \bar c$ (left panel), $D \bar D$ (middle panel) and $e^+ e^-$ 
%(right panel). Here Kwiecinski UGDF and Peterson fragmentation function were
%used.
%}
% \label{fig:p1t_vs_p2t}
%\end{figure}
%-----------------------------------------------------------------------------

If the detector can measure both transverse momenta of electron/positron
and their directions,
one can construct a distribution in transverse momentum of the dielectron
pair: $\vec{p}_{t,sum} = \vec{p}_{1t} + \vec{p}_{2t}$.
Our predictions including the semileptonic decays and Drell-Yan processes
are shown in the left panel of Fig.\ref{fig:new_distributions}. 
Both processes give rather similar distributions. To our knowledge 
the distributions of this type were never measured experimentally.
% as they cannot easily be compared to the calculations in the collinear 
% approach due to its inherent singularities.
%Obviously this is not the case for the $k_t$-factorization approach
%discussed in the present analysis.
The distribution in $p_{t,sum}$ is not only a consequence
of gluon transverse momenta %, as it is for quark and antiquark production,
but invlolves also fragmentation process and semileptonic decays. 
%A measurement of this quantity would test then all stages of the process.
With good azimuthal granulation of detectors one could also construct
distribution in azimuthal angle between electron and positron. 
Corresponding predictions are shown in the right panel of 
Fig.\ref{fig:new_distributions}. One can see 
an interesting dependence on the invariant mass of the dielectron pair
-- the smaller the invariant mass the large the decorrelation in azimuthal
angle.

%--------------------------------------------------------------------------
%\vspace{-2mm}
\begin{figure}[!h]
\begin{center}
\includegraphics[width=5.0cm]{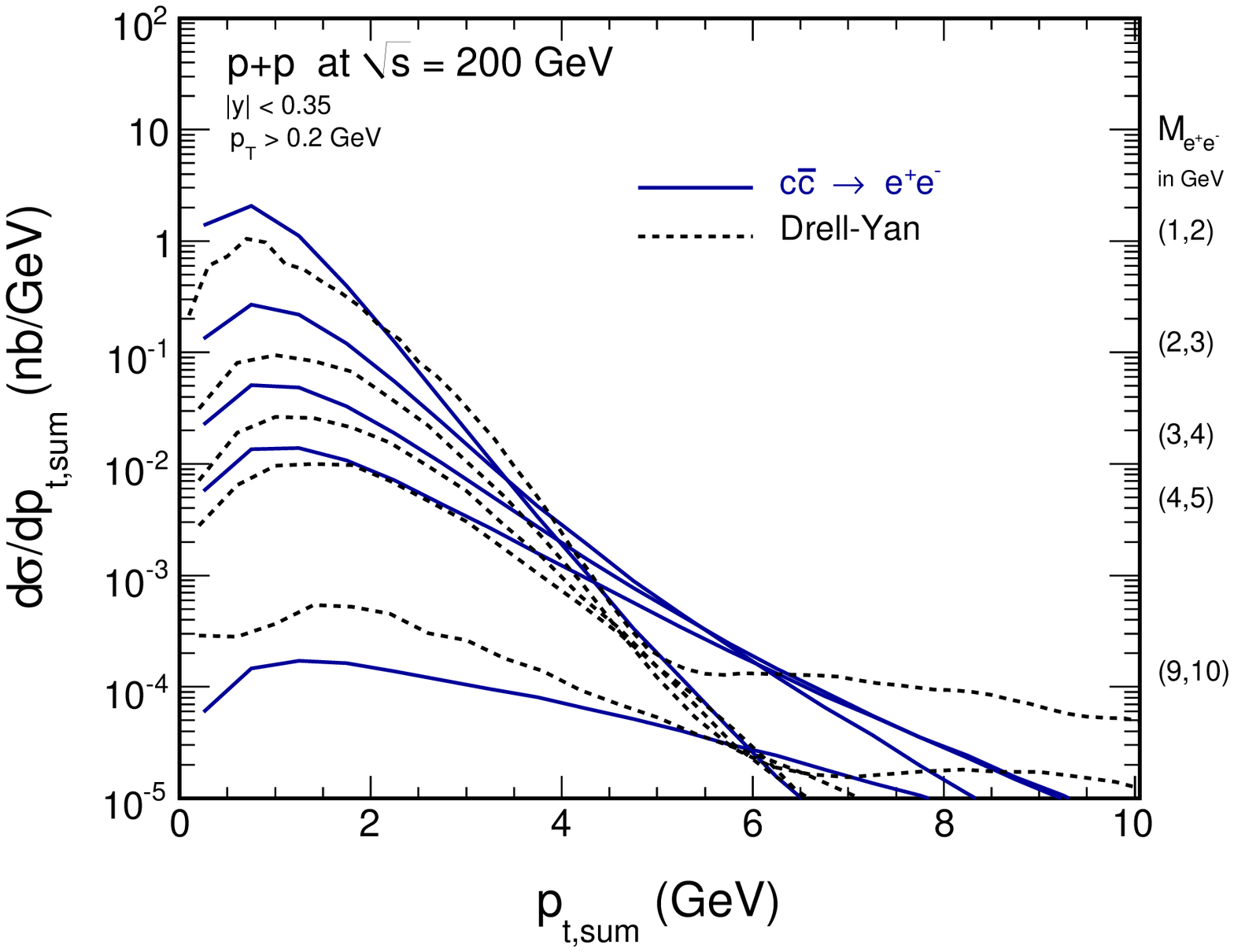}
\includegraphics[width=5.0cm]{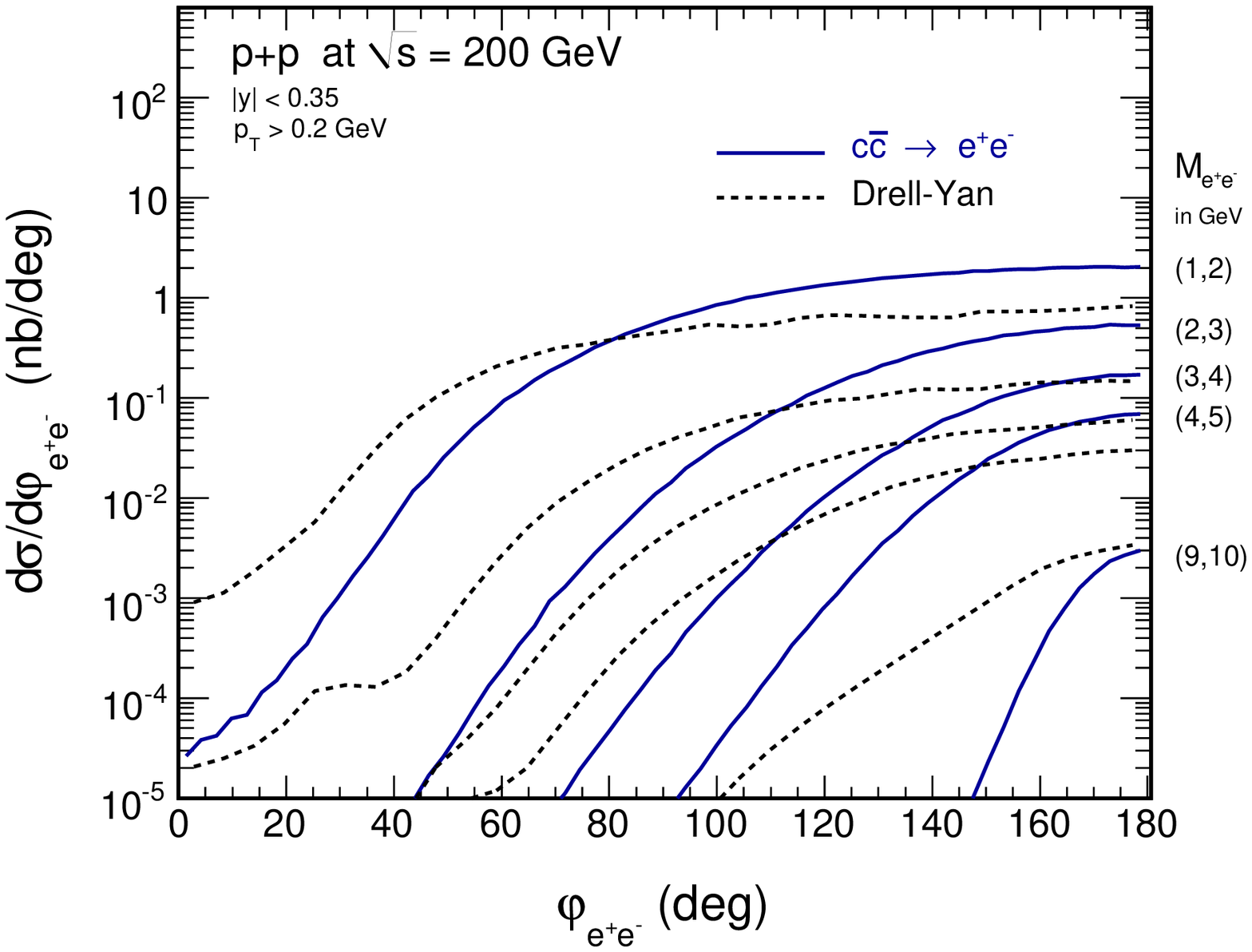}
\end{center}
   \caption{
 \small
Distribution in transverse momentum of the dielectron pair (left)
and in azimuthal angle between electron and positron (right) 
for semileptonic decays (solid line) 
and Drell-Yan processes (dashed line). Here Kwiecinski UGDF and Peterson
fragmentation function were used.
}
 \label{fig:new_distributions}
\end{figure}
%----------------------------------------------------------------------------

%--------------------------------------------------------------------------
%\begin{figure}[!h]
%\begin{center}
%\includegraphics[width=7cm]{fig-azimuth.eps}
%\end{center}
%   \caption{
% \small
%Distribution in azimuthal angle between electron and positron 
%for semileptonic decays (solid line) 
%and Drell-Yan processes (dashed line). Here Kwiecinski UGDFs and Peterson fragmentation functions were
%used.
%}
% \label{fig:azimuth}
%\end{figure}
%-----------------------------------------------------------------------------

%--------------------------
\section{Conclusions}
%--------------------------

We have calculated inclusive spectra of nonphotonic
electrons/positrons for RHIC energy in the framework
of the $k_t$-factorization. We have concentrated on 
the dominant gluon-gluon fusion mechanism and used
unintegrated gluon distribution functions
from the literature.
Special emphasis was devoted
to the Kwieci\'nski unintegrated gluon (parton)
distributions. In this formalism, using unintegrated
quark and antiquark distributions, one can calculate
in addition the quark-antiquark annihilation process
including transverse momenta of initial
quarks/antiquarks. 
%In addition, we have used
%unintegrated gluon distributions constructed by Ivanov and
%Nikolaev to describe deep-inelastic data measured at HERA.

When calculating spectra of charmed ($D$, $D^*$) and 
bottom ($B$, $B^*$) mesons we have used Peterson
and Braaten et al. fragmentation functions.
% with model parameters from the literature. There are small 
%differences between results obtained with both 
%fragmentation functions.
%A very important ingredient, which influences the final
%spectra, is the distribution of
%electrons/positrons from the decay of $D$ and $B$ mesons.
We have used recent fits to the CLEO and BABAR 
collaborations data for decay functions of heavy mesons.
%The  momentum spectra of electrons/positrons
%from the decays of $D$ and $B$ mesons produced in 
%the $e^+ e^-$ collisions were used in the present 
%calculation to generate distribution of 
%electrons/positrons coming from the decays of 
%$D$ and $B$ mesons produced in the hadronic reactions.
%This way we have avoided all uncertainties associated with
%modeling semileptonic decays of mesons.

Our results have been confronted with experimental data measured
recently by the PHENIX and STAR collaborations at RHIC. A reasonable 
description of the data at large transverse momenta of 
electrons/positrons have been achieved. We have discussed
uncertainties related to the choice of the factorization
and renormalization scales and
those related to the fragmentation 
process. Although the uncertainty bands are rather large, 
%similarly as for the higher-order collinear approach, 
there is a missing strength at lower electron/positron transverse momenta.

We have discussed also correlations of charmed mesons
and dielectrons at the energy of recent RHIC experiments. We have
calculated the spectra in dielectron invariant mass, in azimuthal angle 
between electron and positron as well as for the distribution in
transverse momentum of the pair.
The uncertainties due to the choice of UGDFs, choice of the
factorization and renormalization scales, have been discussed.
%choice of the heavy quark
%masses as well as fragmentation functions have been quantified. 
%The uncertainties for UGDFs are larger than those for fragmentation functions.
We have obtained good description of the dielectron invariant mass
distribution measured recently by the PHENIX collaboration at RHIC. 

At RHIC the contribution of electrons from Drell-Yan processes is only 
slightly smaller than that from the semileptonic decays. The
distributions in azimuthal angle between electron and positron and in 
the transverse momentum of the dielectron pair from both processes are 
rather similar. 
%We do not find a possibility of a clear separation of both processes.
It was found that the distribution in azimuthal angle strongly depends
on dielectron invariant mass.

We have also included exclusive central-diffractive contribution
discussed recently in the literature.
At the rather low RHIC energy it gives, however, a very small contribution
to the cross section and can be safely ignored. 

The QED double-elastic, double-inelastic, elastic-inelastic and 
inelastic-elastic processes give individually
rather small contribution but when added together are not negligible
especially at low dielectron invariant masses where some strength is 
missing.

%In the present analysis we have studied correlations between electron
%and positron. It can be also interesting to look at correlations between
%a $D$ meson and electron.
%This will be a subject of a forthcoming analysis.

\vspace{0.5cm}
%--------------------
%\begin{center}
%{\bf Acknowledgments}\\
%\end{center}
%--------------------
%This work was supported in part by the MNiSW grant
%Polish Ministry of Science and Higher Education 
%under grant 
%No. DEC-2011/01/N/ST2/04116.

%==================================================================


\begin{thebibliography}{1000}

\bibitem{STAR_electrons}
J. Adams et al.(STAR collaboration),
Phys. Rev. Lett. {\bf 94} (2005) 062301;\\
B.I. Abelev et al.(STAR collaboration),
Phys. Rev. Lett. {\bf 98} (2007) 192301.

\bibitem{PHENIX_electrons}
A. Adare et al.(PHENIX collaboration),
Phys. Rev. Lett. {\bf 97} (2006) 252002,
 arXiv:hep-ex/0609010.

\bibitem{CLEO}
N.E. Adam, et al. (CLEO collaboration), Phys. Rev. Lett.
{\bf 97} (2006) 251801,
hep-ex/0604044.

\bibitem{BABAR}
B. Aubert, et al. (BABAR collaboration), Phys. Rev. 
{\bf D69} (2004) 111104(R).

\bibitem{Tevatron_mesons}
D. Acosta et al. (CDF II collaboration), Phys. Rev. Lett. {\bf 91} (2003)
241804.

\bibitem{electrons}
A. Adare et al. (PHENIX collaboration), Phys. Rev. Lett. {\bf 97} (2006)
252002.


\bibitem{PHENIX}
A. Adare et al. (PHENIX collaboration), Phys. Lett. {\bf B670}, 313 (2009).


\bibitem{Kwiecinski}
J. Kwieci\'nski, Acta Phys. Polon. {\bf B33} (2002) 1809.
A. Gawron and J. Kwieci\'nski, Acta Phys. Polon. {\bf B34}
(2003) 133.
A. Gawron, J. Kwieci\'nski and W. Broniowski, Phys. Rev. {\bf D68} (2003) 054001.

\bibitem{Peterson}
C. Peterson, D. Schlatter, I. Schmitt, P.M. Zerwas,
Phys. Rev. {\bf D27} (1983) 105.

\bibitem{BCFY95}
E. Braaten, K. Cheung, S. Fleming and T.C. Yuan,
Phys. Rev. {\bf D51} (1995) 4819.

\bibitem{CNV05}
M. Cacciari, P. Nason and R. Vogt, Phys. Rev. Lett. 
{\bf 95} (2005) 122001.

\bibitem{LMS09}
M. {\L}uszczak, R. Maciu{\l}a and A. Szczurek, Phys. Rev. {\bf D79}
(2009) 034009.

\bibitem{MSS11}
R. Maciu{\l}a, A. Szczurek and G. \'Slipek,
%``Kinematical correlations of dielectrons from semileptonic decays
%of heavy mesons and Drell-Yan processes at BNL RHIC'',
Phys. Rev. {\bf D83} (2011) 054014.

\bibitem{LS06}
M. {\L}uszczak and A. Szczurek, Phys. Rev. {\bf D73} (2006) 054028.

%\bibitem{Tevatron_mesons}
%D. Acosta et al. (CDF II collaboration), Phys. Rev. Lett. {\bf 91} (2003)
%241804.

%\bibitem{electrons}
%A. Adare et al. (PHENIX collaboration), Phys. Rev. Lett. {\bf 97} (2006)
%252002.

%\bibitem{FONLL}
%M. Cacciari, M. Greco and P. Nason, JHEP {\bf 9805} (1998) 007;\\
%M. Cacciari, S. Frixione, P. Nason, JHEP {\bf 0103} (2001) 006.

%\bibitem{CNV05}
%M. Cacciari, P. Nason and R. Vogt, Phys. Rev. Lett. 
%{\bf 95} (2005) 122001.

\bibitem{CCH91}
S. Catani, M. Ciafaloni and F. Hautmann, Nucl. Phys. {\bf 366} (1991)
135.

\bibitem{CE91}
J.C. Collins and R.K. Ellis, Nucl. Phys. {\bf B360} (1991) 3.

\bibitem{BE01}
R.D. Ball and R.K. Ellis, J.H.E.P. {\bf 0105} (2001) 053.

%\bibitem{RSS}
%E.M. Levin, M.G. Ryskin, Yu.M. Shabelski and A.G. Shuvaev,
%Sov. J. Nucl. Phys. {\bf 53} (1991) 657; \\
%M.G. Ryskin, Yu.M. Shabelski and A.G. Shuvaev,
%Z. Phys. {\bf C69} (1996) 269; \\
%Yu.M. Shabelski and A.G. Shuvaev, 
%Eur. Phys. J. {\bf C6} (1999) 313;\\
%M.G. Ryskin, A.G. Shuvaev and Yu.M. Shabelski,
%Phys. Atom. Nucl. {\bf 64} (2001) 1995;
%Yu.M. Shabelski and A.G. Shuvaev, hep-ph/0107106;
%Yu.M. Shabelski and A.G. Shuvaev, hep-ph/0406157.

%\bibitem{BS00}
%S.P. Baranov and M. Smizanska, Phys. Rev. {\bf D62} (2000) 014012.

%\bibitem{HKSST00}
%Ph. Hagler, R. Kirschner, A. Sch\"afer, L. Szymanowski
%and O.V. Teryaev, Phys. Rev. {\bf D62} (2000) 071502.

%\bibitem{LSZ02}
%A.V. Lipatov, V.A. Saleev and N.P. Zotov, hep-ph/0112114;\\
%S.P. Baranov, A.V. Lipatov and N.P. Zotov, hep-ph/0302171,
%Yad. Fiz. {\bf 67} (2004) 856.

%\bibitem{LS06}
%M. {\L}uszczak and A. Szczurek, Phys. Rev. {\bf D73} (2006) 054028.

%\bibitem{Peterson}
%C. Peterson, D. Schlatter, I. Schmitt, P.M. Zerwas,
%Phys. Rev. {\bf D27} (1983) 105.

%\bibitem{BCFY95}
%E. Braaten, K. Cheung, S. Fleming and T.C. Yuan,
%Phys. Rev. {\bf D51} (1995) 4819.

%\bibitem{Hill}
%R. J. Hill, arXiv:hep-ph/0606023.

%\bibitem{Mahlke07}
%H. Mahlke, arXiv:hep-ex/0702014.

%\bibitem{AMP08}
%M. Artuso, B. Meadows, A. A. Petrov,
%arXiv:0802.2934.

\bibitem{PDG_new}
C. Amsler et al.[Partilce Data Group], Phys. Lett. 
{\bf B667} (2008) 1. 

%\bibitem{Kwiecinski}
%J. Kwieci\'nski, Acta Phys. Polon. {\bf B33} (2002) 1809.
%A. Gawron and J. Kwieci\'nski, Acta Phys. Polon. {\bf B34}
%(2003) 133.
%A. Gawron, J. Kwieci\'nski and W. Broniowski, Phys. Rev. {\bf D68} (2003) 054001.

%\bibitem{IN02}
%I.P. Ivanov and N.N. Nikolaev, Phys. Rev. {\bf D65} (2002) 054004.

%\bibitem{BP_book}
%V.D. Barger and R.J.N. Phillips, ``Collider physics'',
%Addison-Wesley Publishing Company, 1987

%\bibitem{SS97}
%D.V. Shirkov and I.L. Solovtsov, Phys. Rev. Lett. {\bf 79}
%(1997) 1209.

\bibitem{STAR_B_to_DB}
X. Lin (STAR Collaboration), Jour. Phys. {\bf G34} (2007)
S821;\\
A.G. Knospe, proceeding of the CHARM 2007 workshop, Ithaca,
August 5-8, 2007;\\
X. Lin, a talk at the international conference 
Quark Matter 2008, Jaipur, India, 
February 4-10, 2009.

\bibitem{Mischke08a}
A. Mischke et al. (STAR collaboration),
J. Phys. {\bf G} in press, arXiv:0804.4601;\\
X. Lin et al. (STAR collaboration),
J. Phys. {\bf G34} (2007) S821.

%\bibitem{Mischke08b}
%A. Mischke, arXiv:0807.1309 (hep-ph).

%\bibitem{PDG}
%W.M. Yao et al., [Partilce Data Group], J. Phys. {\bf G33} (2006) 1.

%\bibitem{asymmetry}
%M.I. Adamovich et al. (WA82 Collaboration),
%Phys. Lett. {\bf B306} (1993) 402;\\
%G.A. Alves et al. (E769 Collaboration),
%Phys. Rev. Lett. {\bf 77} (1996) 402;\\
%E.M. Aitala et al. (E791 Collaboration),
%Phys. Lett. {\bf B411} (1997) 230;\\
%M.I. Adamovich et al. (WA92 Collaboration),
%Nucl. Phys. {\bf B495} (1997) 3;\\
%M.I. Adamovich et al. (WA89 Collaboration),
%Eur. Phys. J. {\bf C8} (1999) 593, {\bf C13} (2000) 247;
%M. Iori et al. (SELEX Collaboration),
%Nucl. Phys. B (Proc. Suppl.) {\bf 75} (1999) 16.

%\bibitem{BL06}
%A.V. Berezhnoy, A.K. Likhoded,
%Phys. Atom. Nucl. {\bf 69} (2006) 103.

%\bibitem{CDNN01}
%F. Carvalho, F.O. Duraes, F.S. Navarra and M. Nielsen,
%Phys. Rev. Lett. {\bf 86} (2001) 5434.

%\bibitem{PHENIX_muons}
%S.S. Adler et al. (PHENIX collaboration),
%Phys. Rev. {\bf D76} (2007) 092002, 
%arXiv:hep-ex/0609032.

%\bibitem{subtraction}
%S.S. Adler et al. (PHENIX collaboration),
%Phys. Rev. Lett. {\bf 96} (2006) 032301;\\
%S.S. Adler et al. (PHENIX collaboration),
%Phys. Rev. Lett. {\bf 98} (2007) 172301.

%\bibitem{PHENIX}
%A. Adare et al. (PHENIX collaboration), Phys. Lett. {\bf B670}, 313 (2009).

%\bibitem{Tevatron_mesons}
%D. Acosta et al. (CDF II collaboration), Phys. Rev. Lett. {\bf 91} (2003)
%241804.

%\bibitem{electrons}
%A. Adare et al. (PHENIX collaboration), Phys. Rev. Lett. {\bf 97} (2006)
%252002.

%\bibitem{cacciari}
%M. Cacciari, P. Nason and R. Vogt, Phys. Rev. Lett. {\bf 95} (2005) 122001.

%\bibitem{LMS09}
%M. {\L}uszczak, R. Maciu{\l}a and A. Szczurek, Phys. Rev. {\bf D79}
%(2009) 034009.

%\bibitem{MSS11}
%R. Maciu{\l}a, A. Szczurek and G. \'Slipek,
%``Kinematical correlations of dielectrons from semileptonic decays
%of heavy mesons and Drell-Yan processes at BNL RHIC'',
%Phys. Rev. {\bf D83} (2011) 054014.

%\bibitem{LS06}
%M. {\L}uszczak and A. Szczurek, Phys. Rev. {\bf D73} (2006) 054028.

%\bibitem{mischke}
%A. Mischke, Phys. Lett. {\bf B671}, 361 (2009).

%\bibitem{SS08}
%A. Szczurek and G. \'Slipek, Phys. Rev. {\bf D78} (2008) 114007.

%\bibitem{Wong}
%Ch-Y. Wong and H. Wang, Phys. Rev. {\bf C58} (1998) 376.

%\bibitem{Murgia}
%U. D'Alesio and F. Murgia, Phys. Rev. {\bf D70} (2004) 074009.

%\bibitem{LS2010}
%P. Lebiedowicz and A. Szczurek, Phys. Rev. {\bf D81} (2010) 036003.

%\bibitem{MRST}
%A.D. Martin, et al., Eur. Phys. J. {\bf C39}, 155 (2005). 

%\bibitem{CCH91}
%S. Catani, M. Ciafaloni and F. Hautmann, Nucl. Phys. {\bf 366} (1991) 135.

%\bibitem{Kwiecinski}
%J. Kwieci\'nski, Acta Phys. Polon. {\bf B33} (2002) 1809.

\bibitem{KMRupdf}
M.A. Kimber, A.D. Martin and M. G. Ryskin, Eur. Phys. J. {\bf C12}, 655 (2000)\\
M.A. Kimber, A.D. Martin and M. G. Ryskin, Phys. Rev. {\bf D63} (2001) 114027-1.

%\bibitem{Kutak}
%K. Kutak and A.M. Stasto
%Eur. Phys. J. {\bf C41}, 343 (2005).

%\bibitem{Peterson}
%C. Peterson, et. al., Phys. Rev. {\bf D27} (1983) 105.

%\bibitem{bcfy}
%E. Braaten, et al., Phys. Rev. {\bf D51} (1995) 4819.

%\bibitem{kartvel}
%V.G. Kartvelishvili, et al., Phys. Lett. {\bf B78}, 615 (1978).

%\bibitem{CLEO}
%N.E. Adam, et al. (CLEO collaboration), Phys. Rev. Lett. {\bf 97} (2006) 251801,\\
%B. Aubert, et al. (BABAR collaboration), Phys. Rev. {\bf D69} (2004) 111104(R).

%\bibitem{DZ}
%M. Dress and D. Zeppenfeld, Phys. Rev. {\bf D39} (1989) 2536.

\bibitem{MPS2010} R. Maciula, R. Pasechnik and A. Szczurek, Phys. Lett. B {\bf 685},
165 (2010);

%\bibitem{KMR_Higgs}
%V.A. Khoze, A.D. Martin and M.G. Ryskin, Phys. Lett. B {\bf 401},
%330 (1997);\\
%V.A. Khoze, A.D. Martin and M.G. Ryskin, Eur. Phys. J. C {\bf 23},
%311 (2002).

%\bibitem{KKMR}
%A.B. Kaidalov, V.A. Khoze, A.D. Martin and M.G. Ryskin, Eur. Phys.
%J. C {\bf 33}, 261 (2004).

%\bibitem{KKMR-spin}
%A.B. Kaidalov, V.A. Khoze, A.D. Martin and M.G. Ryskin, Eur.\ Phys.\
%J.\ C {\bf 31}, 387 (2003) [arXiv:hep-ph/0307064].


%\bibitem{KMR_chic}
%L.A. Harland-Lang, V.A. Khoze, M.G. Ryskin and W.J. Stirling,
%Eur.\ Phys.\ J.\ C {\bf 65}, 433 (2010) [arXiv:0909.4748].
%\bibitem{MR}
%  A.~D.~Martin and M.~G.~Ryskin,
%``Unintegrated generalised parton distributions,''
%  Phys.\ Rev.\  D {\bf 64}, 094017 (2001)
%  [arXiv:hep-ph/0107149].
%%CITATION = PHRVA,D64,094017;%%

%\bibitem{GRV95}
%M. Gl\"uck, E. Reya and A. Vogt, Z. Phys. {\bf C67} (1995) 433.

%\bibitem{PDG}
%W.M. Yao, et al. [Particle Data Group], J. Phys. {\bf G33} (2006) 1.


\end{thebibliography}
\end{document}